\documentclass[onecolumn,12pt]{IEEEtran}

\usepackage{amssymb}
\usepackage{graphicx}
\usepackage{amsmath}
\usepackage[english]{babel}
\usepackage{mathrsfs}

\newtheorem{thm}{Theorem}
\newtheorem{theorem}[thm]{Theorem}
\newtheorem{lemma}[thm]{Lemma}

\newtheorem{proposition}[thm]{Proposition}
\newtheorem{corollary}[thm]{Corollary}

\newtheorem{example}{Example}

\newtheorem{remark}{Remark}

\begin{document}

\title {The Shannon Cipher System with a Guessing Wiretapper: General Sources}
\author {Manjesh Kumar Hanawal and Rajesh Sundaresan
\thanks{This work was supported by the Defence Research and Development Organisation,
Ministry of Defence, Government of India, under the DRDO-IISc
Programme on Advanced Research in Mathematical Engineering, and by
the University Grants Commission under Grant Part (2B)
UGC-CAS-(Ph.IV).}
\thanks{The material in this paper was presented in part at the IISc Centenary Conference on Managing Complexity in a Distributed World, (MCDES 2008) held in Bangalore, India, May 2008. A part of this work was presented at the IEEE International Symposium on Information Theory (ISIT 2009) held in Seoul, Korea, June 2009.} }
\maketitle

\begin{abstract}
The Shannon cipher system is studied in the context of general sources using a notion of computational secrecy introduced by Merhav \& Arikan. Bounds are derived on limiting exponents of guessing moments for general sources. The bounds are shown to be tight for iid, Markov, and unifilar sources, thus recovering some known results. A close relationship between error exponents and correct decoding exponents for fixed rate source compression on the one hand and exponents for guessing moments on the other hand is established.
\end{abstract}

\begin{IEEEkeywords}
cipher systems, correct decoding exponent, error exponent, information spectrum, key rate, length function, large deviations, secrecy, sources with memory, fixed-rate source coding
\end{IEEEkeywords}

\section{INTRODUCTION}
\label{sec:introduction}

We consider the classical cipher system of Shannon \cite{194910BSTJ_Sha}. Let $X^n = (X_1, \cdots, X_n)$ be a message
where each letter takes values on a finite set $\mathbb{X}$. This message should be communicated securely from a transmitter to a
receiver, both of which have access to a common secure key $U^k$ of $k$ purely random bits independent of $X^n$. The transmitter
computes the cryptogram $Y = f_n(X^n,U^k)$ and sends it to the receiver over a public channel. The cryptogram may be of variable
length. The encryption function $f_n$ is invertible for any fixed $U^k$. The receiver, knowing $Y$ and $U^k$, computes $X^n = f_n^{-1}(Y,U^k)$. The
functions $f_n$ and $f_n^{-1}$ are published. A wiretapping attacker has access to the cryptogram $Y$, knows $f_n$ and
$f_n^{-1}$, and attempts to identify $X^n$ without knowledge of $U^k$. The attacker can use knowledge of the statistics of $X^n$. We
assume that the attacker has a test mechanism that tells him whether a guess $\hat{X}^n$ is correct or not. For example, the attacker may
wish to attack an encrypted password or personal information to gain access to, say, a computer account, or a bank account via internet,
or a classified database \cite{199909TIT_MerAri}. In these situations, successful entry into the system provides
the natural test mechanism. We assume that the attacker is allowed an unlimited number of guesses.
The {\em key rate} for the cipher system is $R = k (\ln 2) / n$ nats\footnote{We shall mostly use {\em nat} as the unit of
information in this paper by taking natural logarithms. $k (\ln 2) / n$ nats per input symbol is the same as $k/n$ bits
per input symbol.} of secrecy per message (or source) letter.

Merhav \& Arikan \cite{199909TIT_MerAri} studied discrete memoryless sources (DMS) in the above setting and characterized the best
attainable moments of the number of guesses required by an attacker. In particular, they showed that for a DMS with
the governing single letter PMF $P$ on $\mathbb{X}$, the value of the optimal exponent for the $\rho$th moment $(\rho >0)$ is given by
\begin{equation}
\label{eqn:IIDGuessExponent}
E(R, \rho) = \max_{Q} \left\{ \rho \min \{ H(Q), R \} - D(Q \parallel P)  \right\}.
\end{equation}
The maximization is over all PMFs $Q$ on $\mathbb{X}$, $H(Q)$ is the Shannon entropy of $Q$, and $D(Q \parallel P)$ is the
Kullback-Leibler divergence between $Q$ and $P$. They also showed that $E(R,\rho)$ increases linearly in $R$ for $R\leq H(P)$, continues to increase in a concave fashion for $R \in [H(P),H^{'}]$, where $H^{'}$ is a threshold, and is constant for $R>H^{'}$. Unlike the classical equivocation rate analysis, atypical sequences do affect the behavior of $E(R,\rho)$ for $R \in [H(P),H^{'}]$ and perfect secrecy is obtained, i.e., cryptogram is uncorrelated with the message, only for $R>H^{'}>H(P)$. Merhav \& Arikan also determined the best achievable performance based on the probability of a large deviation in the number of guesses, and showed that it equals the Legendre-Fenchel transform of $E(R,\rho)$ as a function of $\rho$. Sundaresan \cite{200805MCDES_Sun} extended the above results to unifilar sources. Hayashi \& Yamamoto \cite{200806TIT_HayYam} proved coding theorems for the Shannon cipher system with correlated outputs $(X^n,Z^n)$ where the wiretapper is interested in $X^n$ while the receiver in $Z^n$.

In this paper, we extend Merhav \& Arikan's notion of  computational secrecy \cite{199909TIT_MerAri} to general sources. One motivation is that secret messages typically come from the natural languages which are modeled well as sources with memory, for e.g., a Markov source of appropriate order. Another motivation is that the study of general sources clearly brings out the connection between guessing and compression, as discussed next.

As with other studies of general sources, {\em information spectrum} plays crucial role in this paper. We show that $E(R, \rho)$ is closely related to (a) the error exponent of a rate-$R$ source code, and (b) the correct decoding exponent of a rate-$R$ source code, when exponentiated probabilities are considered (see Sec. \ref{subsubsec:CorrectDecodingExponent}). In particular, the exponents in (a) and (b) appear in the first and second terms below when we rewrite $E(R,\rho)$ for a DMS as
\begin{eqnarray*}
\lefteqn { E(R,\rho)= \max\bigg\{ \rho R- \min_{Q:H(Q)>R} D(Q\parallel P), }\\
&& \hspace*{1in}\min_{Q:H(Q)\leq R}\{\rho H(Q)-D(Q\parallel P)\} \bigg\}.
\end{eqnarray*}
This brings out the fundamental connection between source coding exponents and  key-rate constrained guessing exponents. Further, unlike the case for the probability of a large deviation in the number of guesses \cite[Sec. V]{199909TIT_MerAri}, both the error exponent and the correct decoding exponent determine $E(R, \rho)$. We extend the above result to general sources by getting upper and lower bounds on $E(R, \rho)$.  We then show that these are tight for DMS, Markov and unifilar sources. The bounds may be of interest even if they are not tight because the upper bound specifies the amount of effort need by an attacker and the lower bound specifies the secrecy strength of the cryptosystem to a designer.

The limiting case as $\rho \downarrow 0$ in (b) yields classical framework for probability of correct decoding. This special case is related to the work of Han \cite{200009TIT_Han} and Iriyama \cite{200105TIT_Iri} who studied the dual problem of rates required to meet a specified error exponent or a specified correct decoding exponent.

The paper is organized as follows. Section \ref{sec:specifiedKeyRateGuessing} relates our problem to a modification of Campbell's compression problem \cite{Campbell-1}. Section \ref{sec:Information Spectrum Results} gives bounds on the limits of exponential rate of guessing moments, in terms of information spectrum quantities. Section \ref{sec:Examples} evaluates the bounds for some specific examples. Section \ref{sec:Conclusion} concludes the paper with additional remarks. Proofs are given in the appendices.

\section{Guessing with key-rate constraints and source compression}
\label{sec:specifiedKeyRateGuessing}

In this section, we make a precise statement of our problem, and establish a connection between guessing and source compression subject to a new cost criterion.

Let $\mathbb{X}^n$ denote the set of messages and $\mathcal{M}(\mathbb{X}^n)$ the set of PMFs on $\mathbb{X}^n$. By a source, we mean a sequence of PMFs $(P_n : n \in \mathbb{N})$, where\footnote{Sometimes we use $P_{X^n}$ in place of $P_n$ when we refer to the distribution of the random vector $X^n$.} $P_n \in \mathcal{M}(\mathbb{X}^n)$. Let $X^n$ denote a message put out by the source and $U^k$ the secure key of $k$ purely random bits independent of $X^n$. Recall that the transmitter computes the cryptogram $Y =f_n(X^n,U^k)$ and sends it to the receiver over a public channel.

For a given cryptogram $Y=y$, define a {\em guessing strategy}
\[
     G_n(\cdot \mid y):~ \mathbb{X}^n   \rightarrow \{1,2,\cdots, |\mathbb{X}|^n\}
\]
as a bijection that denotes the order in which elements of $\mathbb{X}^n$ are guessed. $G_n(x^n\mid y)=l$ indicates that $x^n$ is the $l$th guess, when the cryptogram is $y$. With knowledge of $P_n$, the encryption function $f_n$, and the cryptogram $Y$, the attacker can exhaustively calculate the posterior probabilities of all plaintexts $P_{X^n|Y}(\cdot \mid y)$ given the cryptogram. The attacker's optimal guessing strategy is then to guess in the decreasing order of these posterior probabilities $P_{X^n|Y}(\cdot \mid y)$. Let us denote this optimal attack strategy as $G_{f_n}$. The key rate for the system is $R = k (\ln 2) / n$ nats of secrecy per source letter. Let $(f_n: n \in \mathbb{N})$ denote the sequence of encryption functions, where $\mathbb{N}$ denotes the set of natural numbers. This sequence is known to the attacker. We assume that the attacker employs the aforementioned optimal guessing strategy.

For a given $\rho > 0$, key rate $R>0$, define the normalized guessing exponent
\[
  E_n^g(R, \rho)
  := \sup_{f_n} \frac{1}{n} \ln \mathbb{E} \left[
  G_{f_n}(X^n \mid Y)^{\rho} \right].
\]
The supremum is taken over all encryption functions. Further define performance limits of guessing moments  as in \cite{199909TIT_MerAri}:
\begin{eqnarray}
\label{eqn:LimsupGuessExpo}
E_u^g(R,\rho):=\limsup_{n \rightarrow \infty} E_n^g(R, \rho) \\
\label{eqn:LiminfGuessExpo}
E_l^g(R,\rho):=\liminf_{n \rightarrow \infty} E_n^g(R, \rho).
\end{eqnarray}

We next define the related compression quantities. A length function $L_n:\mathbb{X}^n \rightarrow \mathbb{N}$ is a mapping that satisfies Kraft's inequality:
\[
\sum_{x^n \in \mathbb{X}^n}\exp_2\{-L_n(x^n)\}\leq 1,
\]
where the code alphabet is taken to be binary and $\exp_2\{a\} = 2^{a}$.
(We shall use $\exp$ to denote the inverse of the natural logarithm $\ln$).
Every length function yields an attack strategy with a performance characterized as follows.
\begin{proposition}
\label{prop:key:LengthBasedGuessing} Let $L_n$ be any length function on $\mathbb{X}^n$. There is a guessing list $G_n$ such
that for any encryption function $f_n$, we have\footnote{We reiterate that $R$ is measured in nats.}
\begin{eqnarray*}
  G_n(x^n \mid y) & \leq & 2 \exp_2 \left\{\min \left\{L_n(x^n), nR/(\ln 2)\right\}\right\} \\
                  & = & 2 \exp \left\{\min \left\{L_n(x^n) \ln 2, nR\right\}\right\}.
\end{eqnarray*}
\end{proposition}

\begin{IEEEproof}
We use a technique of Merhav \& Arikan \cite{199909TIT_MerAri}. Let $G_{L_n}$ denote the guessing function that ignores the cryptogram and proceeds in the increasing order of $L_n$ lengths. Suppose $G_{L_n}$ proceeds in the order $x_1^n, x_2^n,
\cdots$. By \cite[Prop. 2]{200812TRPME_HanSun}, we need at most $\exp_2\{L_n(x^n)\}$ guesses to identify $x^n$ (This is a simple consequence of the fact that there are at most $\exp_2\{L_n(x^n)\}$ strings of length less than or equal to $L_n(x^n)$).

\noindent
As an alternative attack, consider the exhaustive key-search attack defined by the following guessing list:
\[
  f_n^{-1}\left( y,u_1^k \right), f_n^{-1}\left( y,u_2^k \right), \cdots
\]
where $u_1^k, u_2^k, \cdots$ is an arbitrary ordering of the keys. This strategy identifies $x^n$ in at most $\exp\{nR\} = \exp_2\{nR/(\ln 2)\}$ guesses.
Finally, let $G_n(\cdot \mid y)$ be the list that alternates between the two lists, skipping those already guessed, i.e., the one that
proceeds in the order
\begin{equation}
  \label{eqn:bestMixedStrategy}
  x_1^n, f_n^{-1}\left( y,u_1^k \right), x_2^n, f_n^{-1}\left( y,u_2^k \right), \cdots .
\end{equation}
Clearly, for every $x^n$, we need at most twice the minimum over the two individual lists.
\end{IEEEproof}

We now look at a weak converse in the expected sense to the above. We first state without proof the following lemma which associates a length function to
any guessing function (see \cite[Prop. 1]{200812TRPME_HanSun}).
\begin{lemma}
\label{lema:GuessLengthRelation}
Given a guessing function $G_n$, there exists a length function $L_{G_n}$ satisfying
\begin{equation}
\label{eqn:GuessLengthRelation}
L_{G_n}(x^n)-1-\log_2 c_n \leq \log_2 G_n(x^n) \leq L_{G_n}(x^n),
\end{equation}
where $$c_n=\sum_{i=1}^{|\mathbb{X}|^n}\frac{1}{i}.$$
\end{lemma}

For a proof, we refer the reader to \cite[Prop. 1]{200812TRPME_HanSun}. We then have the following proposition.
\begin{proposition}
\label{prop:key:OptimalEncryption} Fix $n \in \mathbb{N}$, $\rho >0$. There is an encryption function $f_n$ and a length function $L_n$ such that
every guessing strategy $G_n$ (and in particular $G_{f_n}$) satisfies
\begin{eqnarray*}
  \lefteqn{ \mathbb{E} \left[ G(X^n \mid Y)^{\rho} \right] } \\
  & \geq & \frac{1}{(2 c_n)^{\rho} (2 +
  \rho)} \mathbb{E} \left[\exp \left\{ \rho \min \left\{ L_n\left(X^n\right) \ln 2, nR \right\} \right\}
  \right].
\end{eqnarray*}
\end{proposition}

\begin{IEEEproof}
See Appendix \ref{subsec:ProofOptimalEncryption}. The proof is an extension of Merhav \& Arikan's proof of \cite[Th.1]{199909TIT_MerAri} to sources with memory. The idea is to identify an encryption mechanism that maps messages of roughly equal probability to each other. Our proof also suggests an asymptotically optimal encryption strategy for sources with memory.
\end{IEEEproof}
\begin{remark}
Note that $c_n \leq 1+n \ln |\mathbb{X}|$, so that
\begin{equation}
  \label{eqn: c_n/n-bound}
  \frac{\log_2 c_n}{n} = O \left( \frac{\log_2 n}{n} \right) = o(1),
\end{equation}
a fact that will be put to good use in the sequel. \hfill \IEEEQEDopen
\end{remark}

Propositions \ref{prop:key:LengthBasedGuessing} and \ref{prop:key:OptimalEncryption} naturally suggest the following
coding problem: identify
\begin{equation}
  \label{eqn:newCodingProblem}
  E_n^s(R, \rho) :=  \min_{L_n} \frac{1}{n} \ln \mathbb{E} \left[ \exp\left\{ \rho \min \left\{ L_n(X^n) \ln 2, nR \right\} \right\} \right].
\end{equation}
The minimum is taken over all length functions. We may interpret the cost of using length $L_n(x^n)$ as $\exp \left \{\min\{L_n(x^n) \ln 2, nR\}\right \}$, i.e., the cost is exponential in $L_n$, but saturates at $\exp\{nR\}$ and so all lengths larger than $nR$ nats (i.e., $nR / (\ln 2)$ bits) enjoy the saturated cost.
Then $E_n^s(R, \rho)$ is the minimum normalized exponent of the $\rho$th moment of this new compression cost.
In analogy with (\ref{eqn:LimsupGuessExpo}) and (\ref{eqn:LiminfGuessExpo}) we define
\[
E_u^s(R, \rho)=\displaystyle \limsup_{n \rightarrow \infty}E_n^s(R, \rho)
\]
\[
E_l^s(R, \rho)=\displaystyle \liminf_{n \rightarrow \infty}E_n^s(R, \rho)
\]
The following is a corollary to Propositions \ref{prop:key:LengthBasedGuessing} and \ref{prop:key:OptimalEncryption}, and relates $E_n^g(R, \rho)$ and $E_n^s(R, \rho)$.
\begin{corollary}
\label{cor:EnEnl} For a given $R, \rho > 0$, we have
\begin{equation}
\label{eqn:DiffEnEnl}
  \left| E_n^s(R,\rho) - E_n^g(R,\rho) \right| \leq \frac{\ln ((4 c_n)^{\rho} (2+\rho))}{n}.
\end{equation}
\end{corollary}

\begin{IEEEproof}
Let $L_n^*$ be the length function that achieves $E_n^s(R,\rho)$. Using Proposition \ref{prop:key:LengthBasedGuessing}, and after taking
expectation, we have the guessing strategy $G_n$ that satisfies
\begin{eqnarray*}
  \lefteqn{  \mathbb{E} \left[ \exp\left\{\rho \min \left\{ L_n^*(X^n) \ln 2, nR\right\} \right\} \right] } \\
  & \geq & \sup_{f_n} \frac{1}{2^{\rho}} \mathbb{E} \left[ G_n(X^n \mid Y)^{\rho}
  \right] \\
  & \geq & \sup_{f_n} \frac{1}{2^{\rho}} \mathbb{E} \left[ G_{f_n}(X^n \mid Y)^{\rho}
  \right] \\
  & \geq & \frac{1}{(4 c_n)^{\rho} (2 + \rho)} \mathbb{E} \left[\exp \left\{\rho \min \left\{ L_n(X^n) \ln 2, nR\right\} \right\} \right] \\
  &&\mbox{for some $f_n$ and $L_n$, given by Proposition \ref{prop:key:OptimalEncryption},} \\
 &\geq& \frac{1}{(4 c_n)^{\rho} (2 + \rho)} \mathbb{E} \left[ \exp \left\{\rho \min \left\{L_n^*(X^n) \ln 2, nR \right\} \right\} \right].
\end{eqnarray*}
Take logarithms, normalize by $n$, use $c_n > 1$ and $\rho > 0$ to get (\ref{eqn:DiffEnEnl}).
\end{IEEEproof}
We now state the equivalence between compression and guessing.
\begin{theorem}[Guessing-Compression Equivalence]
\label{prop:guessing-sourcecode-equivalence}
For any $\rho >0$ and $R>0$, we have $E_u^s(R, \rho)=E_u^g(R,\rho)$ and $E_l^s(R, \rho)=E_l^g(R,\rho)$.
\end{theorem}

\begin{IEEEproof}\
From Corollary \ref{cor:EnEnl} and (\ref{eqn: c_n/n-bound}), magnitude of the difference between $E_n^g(R, \rho)$ and $E_n^s(R, \rho)$ decays as $O((\ln n)/n)$ and vanishes as $n \rightarrow \infty$.
\end{IEEEproof}

Thus, the problem of finding the optimal guessing exponent is the same as that of finding the optimal exponent for the coding problem in (\ref{eqn:newCodingProblem}). When $R \geq\ln |\mathbb{X}|$, the coding problem in (\ref{eqn:newCodingProblem}) reduces to the one considered by Campbell in \cite{Campbell-1}; this is a case where perfect secrecy is obtained and is studied in \cite{200812TRPME_HanSun}. Proposition \ref{prop:key:LengthBasedGuessing} shows that the optimal length function attaining the minimum in (\ref{eqn:newCodingProblem}) yields an asymptotically optimal attack strategy on the cipher system. Moreover, the encryption strategy in the proof of Proposition \ref{prop:key:OptimalEncryption} (see Appendix \ref{subsec:ProofOptimalEncryption}) is asymptotically optimal, from the designer's point of view.

In the rest of the paper we focus on the equivalent compression problem and find bounds on $E_u^s$ and $E_l^s$.

\section{Growth Exponent for the Modified Compression Problem}
\label{sec:Information Spectrum Results}

We begin with some words on notation. Recall that $\mathcal{M}(\mathbb{X}^n)$ denotes the set of PMFs on $\mathbb{X}^n$. The Shannon entropy for a $P_n \in \mathcal{M}(\mathbb{X}^n)$ is
\[
  H(P_n)=-\sum_{x^n \in \mathbb{X}^n} P_n(x^n) \ln P_n(x^n)
\]
and the R\'{e}nyi entropy of order $\alpha \neq 1$ is
\begin{equation}
  \label{eqn:RE}
  H_{\alpha}(P_n)=\frac{1}{1-\alpha} \ln \left( \sum_{x^n \in \mathbb{X}^n} P_n(x^n)^{\alpha} \right).
\end{equation}
The Kullback-Leibler divergence or relative entropy between two PMFs $Q_n$ and $P_n$ is
\[
  D(Q_n \parallel P_n) = \left\{
                       \begin{array}{ll}
                         \hspace*{-.1in} \displaystyle \sum_{x^n \in \mathbb{X}^n }Q_n(x^n) \ln  \frac{Q_n(x^n)}{P_n(x^n)}, & \mbox{if } Q_n \ll P_n, \\
                         \infty, & \mbox{otherwise,}
                       \end{array}
  \right.
\]
where $Q_n \ll P_n$ means $Q_n$ is absolutely continuous with respect to $P_n$. We shall use $\left(X^n: n \in \mathbb{N}\right)$ to denote a sequence of random variables on $\mathbb{X}^n$, with corresponding sequence of probability measures denoted by $\mathbf{X}:=\left(P_{X^n}: n \in  \mathbb{N}\right)$. Thus $\mathbf{X}$ is a source and $X^n$ its $n$-letter message output. Abusing notation, we let $\mathcal{M}(\mathbb{X}^{\mathbb{N}})$  denote the set of all sequences $\mathbf{Y}=\left(P_{Y^n}: n \in  \mathbb{N}\right)$ of probability measures, and for each $\mathbf{B}:=\left(B_n \subseteq \mathbb{X}^n:n \in  \mathbb{N}\right)$, we define
\[
\mathcal{M}(\mathbf{B}):=\left\{\mathbf{Y} \in \mathcal{M}(\mathbb{X}):\displaystyle \lim_{n \rightarrow \infty}P_{Y^n}(B_n)=1\right \}.
\] In the rest of this section $\mathbf{X}$ is a fixed source.
For any $\mathbf{Y} \in \mathcal{M}(\mathbf{B})$ and $\rho > 0$, define
\begin{equation*}
E_u(\mathbf{Y},\mathbf{X},\rho):=\displaystyle \limsup_{n \rightarrow \infty}\frac{1}{n}\{\rho H(P_{Y^n})-D(P_{Y^n}\parallel P_{X^n})\}
\end{equation*}
and
\begin{equation*}
E_l(\mathbf{Y},\mathbf{X},\rho):=\liminf_{n \rightarrow \infty}\frac{1}{n}\{\rho H(P_{Y^n})-D(P_{Y^n}\parallel P_{X^n})\}.
\end{equation*}

We next state a large deviation result that plays a key role in the derivation of bounds on $E_u^s$ and $E_l^s$.

\begin{proposition}
\label{prop: limit-variational-formula}
For all $\rho \geq 0$ and $\mathbf{B}=(B_n\subseteq \mathbb{X}^n:n \in \mathbb{N})$, we have
\begin{equation}
\label{eqn:limsup-Eu}
(1+\rho)\limsup_{n \rightarrow \infty}\frac{1}{n}\ln \displaystyle \sum_{x^n \in B_n}P_{X^n}^{\frac{1}{1+\rho}}(x^n)=\max_{\mathbf{Y} \in \mathcal{M}(\mathbf{B})}E_u(\mathbf{Y},\mathbf{X},\rho)
\end{equation}
\begin{equation}
\label{eqn:liminf-El}
(1+\rho)\liminf_{n \rightarrow \infty}\frac{1}{n}\ln \sum_{x^n \in B_n}P_{X^n}^{\frac{1}{1+\rho}}(x^n)=\max_{\mathbf{Y} \in \mathcal{M}(\mathbf{B})}E_l(\mathbf{Y},\mathbf{X},\rho)
\end{equation}
The maximum-achieving distribution in (\ref{eqn:limsup-Eu}) and (\ref{eqn:liminf-El}) is the source $\mathbf{X}^*=(P_{X^n}^* : n \in \mathbb{N})$ given by
\begin{equation}
\label{eqn:macachiev-sequnce-distribution}
P_{X^n}^*(\cdot)=\frac{P_{X^n}^{\frac{1}{1+\rho}}(\cdot)}{\sum_{y^n \in B_n}P_{X^n}^{\frac{1}{1+\rho}}(y^n)}.
\end{equation}
\end{proposition}

\begin{IEEEproof}
See Appendix \ref{subsec:limit-variational-formula-proof}.
\end{IEEEproof}
\noindent
\begin{remark}
This proposition is a generalization of Iriyama's \cite[Prop. 1]{200105TIT_Iri}, which is obtained by setting $\rho=0$. \hfill \IEEEQEDopen
\end{remark}

\subsection{Upper Bound on $E_u^s$}
\label{subsec:UpperBound}
We first obtain an upper bound on $E_u^s$. We use $\mathbb{E}_{X^n}[\cdot]$ to denote  the expectation with respect to distribution $P_{X^n}$.

\begin{proposition}[Upper Bound]
\label{prop:UpperBound}
Let $R >0$ and $\rho > 0$. Then
\[E_u^s(R,\rho) \leq  \min_{ 0\leq \theta\leq \rho}    \left [(\rho -\theta)R + \max_{\mathbf{Y} \in \mathcal{M}(\mathbb{X}^\mathbb{N})}E_u(\mathbf{Y},\mathbf{X},\theta)\right ]. \]
\end{proposition}

\begin{IEEEproof}
We first recall the useful variational formula \cite[Prop. 1.4.2]{1999AWCATLDP_DupEll}
\begin{eqnarray}
\lefteqn {\ln \mathbb{E}_{X^n}\left [\exp\{U(X^n)\}\right]} \nonumber\\
\label{eqn:VariationalEquation}
&&= \sup_{P_{Y^n}}\left\{\mathbb{E}_{Y^n}[U(Y^n)]-D(P_{Y^n} \parallel P_{X^n})\right\}
\end{eqnarray}
for any $U: \mathbb{X}^n \rightarrow \mathbb{R}$, where $\mathbb{R}$ denotes set of real numbers. For notational convenience, let $d({Y^n}) :=D(P_{Y^n}\parallel P_{X^n})$.
Observe that
\begin{eqnarray}
\lefteqn {\ln \mathbb{E}_{X^n} \left[ \exp\left \{\rho \min\{L_n(X^n) \ln 2, nR\}\right \}  \right ]} \nonumber \\
\label{eqn:vari_formula_ubound}
&=& \sup_{P_{Y^n}} [ \rho \mathbb{E}_{Y^n} \left [\min\{L_n(Y^n) \ln 2, nR\} \right ] - ~ d({Y^n}) ]  \\
\label{eqn:Jesen-applic-1}
&\leq&  \sup_{P_{Y^n}} [ \rho \min\{\mathbb{E}_{Y^n} \left [L_n(Y^n) \ln 2\right ], nR\} - ~ d({Y^n}) ]  \\
&=& \sup_{P_{Y^n}} \bigg \{  \min_{ 0\leq \theta\leq \rho}  [(\rho -\theta)nR +\theta \mathbb{E}_{Y^n} \left [L_n(Y^n) \ln 2\right ] \nonumber \\
\label{eqn:min-linear-combo}
& & \hspace*{2in} - ~ d({Y^n}) \bigg \} \\
&=& \min_{ 0\leq \theta\leq \rho}  \sup_{P_{Y^n}} \bigg \{(\rho -\theta)nR +\theta \mathbb{E}_{Y^n} \left [L_n(Y^n) \ln 2\right ] \nonumber \\
\label{eqn:supinf-interchange-first}
& & \hspace*{2in} - ~ d({Y^n})\bigg \} \\
&=& \min_{ 0\leq \theta\leq \rho}   \bigg\{(\rho -\theta)nR +\sup_{P_{Y^n}}\bigg \{\theta \mathbb{E}_{Y^n}\left [L_n(Y^n) \ln 2\right ] \nonumber \\
& & \hspace*{2in} - ~ d({Y^n}) \bigg\} \bigg\}. \nonumber
\end{eqnarray}
In the above sequence of inequalities, (\ref{eqn:vari_formula_ubound}) follows from  the variational formula (\ref{eqn:VariationalEquation}) with $$U(x^n)=\rho\min\{L_n(x^n) \ln 2, nR\}.$$ Inequality (\ref{eqn:Jesen-applic-1}) follows from Jensen's inequality because $\min\{\cdot,nR\} $ is concave for a fixed $nR$. Equality (\ref{eqn:min-linear-combo}) follows from the identity
\[\rho\min\{a,b\}=\min_{0\leq \theta \leq \rho}\{\theta a + (\rho-\theta) b\}.\]
Equality (\ref{eqn:supinf-interchange-first}) follows because the term within braces is linear in $\theta$ for a fixed $P_{Y^n}$, concave in $P_{Y^n}$ for a fixed $\theta$, and the sets $[0, \rho]$ and $\mathcal{M}(\mathbb{X}^n)$ are compact and convex; these permit an interchange of sup and inf, thanks to a minmax theorem
\cite[Cor. 2, p. 53]{1954xxTGSD_BlaGir}.
Taking $\inf$ over $L_n$, and interchanging the $\inf$ over $L_n$ and the $\min$ over $\theta$, we get
\begin{eqnarray}
\lefteqn {\inf_{L_n}\ln \mathbb{E}_{X^n} \left[ \exp\left \{\rho \min\{L_n(Y^n) \ln 2, nR\}\right \}  \right ]} \nonumber\\
&\leq& \min_{ 0\leq \theta\leq \rho}    \bigg \{(\rho -\theta)nR + \inf_{L_n} \sup_{P_{Y^n}}\bigg \{\theta \mathbb{E}_{Y^n} \left [L_n(Y^n) \ln 2\right ] \nonumber \\
&&\hspace*{2.1in}  - ~ d({Y^n})\bigg \}\bigg \}\nonumber\\
&=& \min_{ 0\leq \theta\leq \rho}    \bigg \{(\rho -\theta)nR + \sup_{P_{Y^n}} \bigg \{\theta \inf_{L_n} \mathbb{E}_{Y^n} \left [L_n(Y^n) \ln 2\right ]  \nonumber\\
\label{eqn:supinf-interchange-second}
&& \hspace{1.5in} - ~ d({Y^n})\bigg \} + O(1)\bigg\} \\
\label{eqn:sourcecodind-entropy-ubound}
&=&  \min_{ 0\leq \theta\leq \rho}    \bigg \{(\rho -\theta)nR + \sup_{P_{Y^n}} \bigg \{\theta H(P_{Y^n})   \nonumber\\
&& \hspace{1.5in} - ~ d({Y^n})\bigg \}+ O(1)\bigg \} \\
\label{eqn:VariationalUBound}
&= & \min_{ 0\leq \theta\leq \rho}    \bigg \{(\rho -\theta)nR + \theta H_{\frac{1}{1+\theta}}(P_{X^n}) + O(1) \bigg \}.
\end{eqnarray}
Equality (\ref{eqn:supinf-interchange-second}) follows because the function inside the inner braces is concave in $P_{Y^n}$, asymptotically linear in $L_n$ (see proof of \cite[Prop. 6]{200812TRPME_HanSun}), and $\mathcal{M}(\mathbb{X}^n)$ is compact; this allows us to interchange $\inf$ and $\sup$. Inequality (\ref{eqn:sourcecodind-entropy-ubound}) follows because $\inf$ of expected compression lengths over all prefix codes is within $\ln 2$ nats (1 bit) of entropy. The last equality follows from the well-known variational characterization of R\'{e}nyi entropy,
\begin{equation}
\label{eqn:RenyiVariationalCharac}
\sup_{P_{Y^n}}\left\{\theta H(P_{Y^n})-D(P_{Y^n}\parallel P_{X^n})\right\}= \theta H_{\frac{1}{1+\theta}}(P_{X^n}),
\end{equation}
a fact that can also be gleaned from the variational formula (\ref{eqn:VariationalEquation}).
Divide both sides of (\ref{eqn:VariationalUBound}) by $n$ and take limit supremum as $n \rightarrow \infty$ to get

\begin{eqnarray}
\lefteqn {E_u^s(R,\rho)} \nonumber\\
&\leq& \limsup_{n \rightarrow \infty }\min_{ 0\leq \theta\leq \rho}    \left \{(\rho -\theta)R +\frac{\theta}{n} H_{\frac{1}{1+\theta}}(P_{X^n}) \right \}  \nonumber \\
&\leq& \min_{ 0\leq \theta\leq \rho}    \left \{(\rho -\theta)R +\theta\limsup_{n \rightarrow \infty }\frac{1}{n} H_{\frac{1}{1+\theta}}(P_{X^n}) \right \}  \nonumber \\
\label{eqn: apply-asyptotic-vartional-inequality}
&=& \min_{ 0\leq \theta\leq \rho}    \left \{(\rho -\theta)R + \max_{\mathbf{Y} \in \mathcal{M}(\mathbb{X}^\mathbb{N})}E_u(\mathbf{Y},\mathbf{X},\theta)\right \}, \nonumber
\end{eqnarray}
where the last inequality follows from Proposition \ref{prop: limit-variational-formula} and the formula for R\'{e}nyi entropy. This completes the proof.
\end{IEEEproof}

From the above proof it is clear that the upper bound holds with equality, when Jensen's inequality holds with equality in (\ref{eqn:Jesen-applic-1}), i.e, the random variable $(1/n)\min\{L_n(X^n) \ln 2, nR\}$ tends asymptotically to a constant. This would happen, for example, when normalized encoded lengths concentrate around the entropy rate of the source.

\subsection{ Lower Bound on $E_l^s$}
\label{subsec:LowerBound}
We now derive a lower bound on $E_l^s$. For a given distribution $P_{Y^n}$ arrange the elements of set $\mathbb{X}^n$ in the decreasing order of their $P_{Y^n}$-probabilities as done in Sundaresan \cite[Sec. IV]{200805MCDES_Sun}. Enumerate the sequences from 1 to $\left |\mathbb{X}\right |^n$. Henceforth refer to a message by its index. Let  $T_R(Y^n)$ denote the first $M=\lfloor\exp\{nR\}\rfloor$ elements in the list. We denote the probability of this set by  $F_{Y^n}$, i.e., \[F_{Y^n}=\sum_{x^n \in T_R(Y^n)} P_{Y^n}(x^n),\] and the probability of the complement of this set $T_R^c(Y^n)$ by $F_{Y^n}^{c}$. Let the  restriction of $P_{Y^n}$ to this set  $T_R(Y^n)$ be $P_{Y^n}^{\prime}$. Let $L_n^*$ denote the length function that attains $E_n^s(R,\rho)$ in (\ref{eqn:newCodingProblem}). As the length functions are uniquely decipherable we have $\exp_2\{L_n^*(i)\}\geq i$.  \\
\begin{proposition}[Lower Bound]
\label{prop:LowerBound}
For a given $\rho > 0$ and rate $R>0$, we have
\begin{eqnarray}
\label{eqn:AymptoticLowerBound}
\lefteqn{ E_l^s(R,\rho) \geq \max \bigg \{\rho R+\liminf_{n \rightarrow \infty}\frac{1}{n}\ln  F_{X^n}^{c} , }\nonumber \\
& & \hspace*{.2in}(1 + \rho)\liminf_{n \rightarrow \infty}\frac{1}{n}\ln \sum_{x^n \in T_R(X^n)}P_{X^n}^{\frac{1}{1+\rho}}(x^n)  \bigg \}.
\end{eqnarray}
\end{proposition}

\begin{remark}
The first term contains limit infimum of the error exponent for a rate-$R$ source code. The second exponent is the correct decoding exponent for a rate-$R$ code when $\rho \downarrow 0$. \hfill \IEEEQEDopen
\end{remark}
\begin{IEEEproof}
The variational formula (\ref{eqn:VariationalEquation}) applied to the function $U(x^n)=\rho \min\{L_n(x^n) \ln 2, nR\}$ gives
\begin{eqnarray}
\lefteqn {\inf_{L_n}\ln \mathbb{E}_{X^n} \left[\exp \left\{\rho \min \left \{ L_n(X^n) \ln 2, nR \right \} \right \} \right] } \nonumber\\
&=& \inf_{L_n}\sup_{P_{Y^n} } \{\rho \mathbb{E}_{Y^n}[\min\{L_n(Y^n) \ln 2, nR\}] -d(Y^n) \} \nonumber \\
&\geq&  \sup_{P_{Y^n} }\bigg \{\rho \inf_{L_n} \mathbb{E}_{Y^n}\left[\min\{L_n(X^n) \ln 2, nR\}\right]-d(Y^n)\bigg \} \nonumber \\
\label{eqn:MaxMinLowerBound}
& &
\end{eqnarray}
where the interchange of inf and sup yields the lower bound in (\ref{eqn:MaxMinLowerBound}). Fix a distribution $P_{Y^n}$ and consider the first term in  (\ref{eqn:MaxMinLowerBound}). Using the enumeration indicated above, we may write
\begin{eqnarray}
\lefteqn {\inf_{L_n} \mathbb{E}_{Y^n}\left[\min\{L_n(Y^n) \ln 2, nR\} \right ]} \nonumber\\
&=& \sum_{i=1}^{|\mathbb{X}|^n} P_{Y^n}(i)\min\{L_n^*(i) \ln 2, nR\} \nonumber\\
&=& \sum_{i=1}^M P_{Y^n}(i)\min\{L_n^*(i) \ln 2, nR\} + \sum_{i=M+1}^{|\mathbb{X}|^n} P_{Y^n}(i)nR \nonumber\\
\label{eqn:optimal Guess}
&\geq&  \sum_{i=1}^{M} P_{Y^n}(i)\ln G_n^*(i) + nRF_{Y^n}^{c} \\
&\geq& F_{Y^n} \sum_{i=1}^M  \frac{P_{Y^n}(i)}{F_{Y^n}}  {L_{G_n^*}(i)} \ln 2 - \ln 2 - \ln (1 + n \ln |\mathbb{X}|) \nonumber \\
\label{eqn:Guess_Lentgh Prop}
& & ~ + ~ nRF_{Y^n}^{c} \\
\label{eqn:lbound-min-expect}
&\geq& F_{Y^n} H(P_{Y^n}^{\prime})- \ln 2 - \ln (1+ n\ln |\mathbb{X}|) + nRF_{Y^n}^{c}.
\end{eqnarray}
Inequality (\ref{eqn:optimal Guess}) follows because
$$L_n^*(i) \ln 2 \geq \ln i = \ln G_n^*(i)$$
with  $G_n^*$ the guessing strategy that guesses in
decreasing order of $P_{Y^n}$ probabilities. $L_{G_n^*}$ in (\ref{eqn:Guess_Lentgh Prop}) denotes the length function given by Lemma \ref{lema:GuessLengthRelation}. Inequality (\ref{eqn:lbound-min-expect}) follows from the source coding theorem's lower bound. Substitute (\ref{eqn:lbound-min-expect}) in (\ref{eqn:MaxMinLowerBound}), normalize by $n$, and take limit infimum to get
\begin{eqnarray*}
\label{eqn:Guess-lowerbound-Inter}
\lefteqn {\displaystyle E_l^s(R,\rho)} \nonumber\\
&\geq& \liminf_{n \rightarrow \infty}\frac{1}{n}\sup_{P_{Y^n}} \bigg \{ \rho F_{Y^n}H(P_{Y^n}^{\prime})+ F_{Y^n}^{c} \rho n R  - d(Y^n) \bigg \}.
\end{eqnarray*}
$P_{Y^n}$ may be thought of as a triplet made of $P_{Y^n}^{\prime},F_{Y^n},$ and the restriction of $P_{Y^n}$ to $T_R^c(Y^n)$. We now perform the optimization
\begin{equation}
\label{eqn:EquationToOptimize}
\sup_{P_{Y^n}}\left \{ \rho F_{Y^n}H(P_{Y^n}^{\prime})+ F_{Y^n}^{c} \rho n R - d(Y^n)\right \}
\end{equation}
in four steps.

\noindent {\bf Step 1}: We first optimize over permutations of probabilities over strings. $F_{Y^n}$, $F_{Y^n}^c$, $H(P_{Y^n})$, and $H(P'_{Y^n})$ remain unchanged over these permutations. Observe that
$$-d(Y^n) = H(P_{Y^n}) + \sum_{y^n} P_{Y^n}(y^n) \ln P_{X^n}(y^n),$$
and so the maximum for $-d(Y^n)$ is attained when the permutation that orders $P_{X^n}(\cdot)$ in decreasing order also orders $P_{Y^n}(\cdot)$ in decreasing order. In particular, $T_R(Y^n)$ equals $T_R(X^n)$.

\noindent {\bf Step 2}: We now optimize over restriction of $P_{Y^n}$ to $T_R^c(Y^n)$. For a fixed $F_{Y^n}$, the log-sum inequality yields
\begin{equation*}
\sum_{x^n \in T_R^c(X^n)}P_{Y^n}(x^n)\ln \frac{P_{Y^n}(x^n)}{P_{X^n}(x^n)} \geq F_{Y^n}^{c} \ln \frac{F_{Y^n}^{c}}{F_{X^n}^{c}},
\end{equation*}
with equality  if and only if $P_{Y^n}(x^n)=P_{X^n}(x^n)\frac{F_{Y^n}^{c}}{F_{X^n}^{c}}$ for all $x^n \in T_R^c(P_{X^n}) $.

\noindent {\bf Step 3}: To optimize over $P_{Y^n}^{\prime}$ rewrite (\ref{eqn:EquationToOptimize}) as
\begin{eqnarray}
\lefteqn {\sup_{P_{Y^n}}\bigg \{\rho F_{Y^n}H(P_{Y^n}^{\prime}) + F_{Y^n}^{c}\rho nR } \nonumber\\
&&\hspace{-.3in}-\sum_{i=1}^{M}P_{Y^n}(i)\ln \frac{P_{Y^n}(i)}{P_{X^n}(i)} -\sum_{M+1}^{\mathbb{|X|}^n}P_{Y^n}(i)\ln \frac{P_{Y^n}(i)}{P_{X^n}(i)}\bigg \} \nonumber\\
\label{eqn: Diver_Lower_Bound}
&=&\sup_{P_{Y^n}^{\prime},F_{Y^n}}\bigg \{\rho F_{Y^n}H(P_{Y^n}^{\prime}) + F_{Y^n}^{c}\rho nR \nonumber\\
&&-\sum_{i=1}^{M} P_{Y^n}(i)\ln \frac{P_{Y^n}(i)}{P_{X^n}(i)} -F_{Y^n}^{c} \ln \frac{F_{Y^n}^{c}}{F_{X^n}^{c}} \bigg \} \\
\label{eqn: Binary_Restricted_Diver}
&=& \sup_{P_{Y^n}^{\prime},F_{Y^n}}\bigg \{\rho F_{Y^n}H(P_{Y^n}^{\prime}) \nonumber + F_{Y^n}^{c} \rho n R \nonumber\\
&&\hspace*{.1in}-F_{Y^n}D(P_{Y^n}^{\prime}\parallel P_{X^n}^{\prime}) - D(F_{Y^n}||F_{X^n}) \bigg \} \\
\label{eqn:Renyi_Entropy}
&=& \sup_{F_{Y^n}}\bigg \{\rho F_{Y^n} H_{\frac{1}{1+\rho}}(P_{X^n}^{\prime}) + F_{Y^n}^{c}\rho nR  \nonumber\\
&& \hspace*{1.4in}-D(F_{Y^n}\parallel F_{X^n}) \bigg \}.
\end{eqnarray}
Equality (\ref{eqn: Diver_Lower_Bound}) is obtained by substituting the attained lower bound in Step~2. In (\ref{eqn: Binary_Restricted_Diver}), $P_{Y^n}^{\prime}$ and $P_{X^n}^{\prime}$ denote conditional distributions of $P_{Y^n}$ and $P_{X^n}$ given $T_R(Y^n)$ and $T_R(X^n)$, respectively, where $T_R(Y^n) = T_R(X^n)$ as argued in Step~1. $D(F_{Y^n}||F_{X^n})$ denotes the divergence between binary random variables  whose probabilities are $\{F_{Y^n},1-F_{Y^n}\}$ and  $\{F_{X^n},1-F_{X^n}\}$ respectively. Finally we used variational characterization of R\'{e}nyi entropy given in (\ref{eqn:RenyiVariationalCharac}) to arrive at (\ref{eqn:Renyi_Entropy}).

\noindent {\bf Step 4}:
We now optimize over $F_{Y^n}\in [0,1]$. Let $Z$ be a binary random variable defined as
\[
  Z = \left\{
                       \begin{array}{ll}
                         \rho H_{\frac{1}{1+\rho}}(P_{X^n}^{\prime}) & \mbox{with probability } ~ F_{Y^n}, \\
                         \rho nR & \mbox{with probability}~ 1-F_{Y^n}
                       \end{array}
  \right.
\]
By $\mathbb{E}_{F_{Y^n}}[Z]$ we mean the expectation of $Z$ with respect to the above distribution. Since $Z$ is a positive random variable, the variational formula yields
\[
  \sup_{F_{Y^n}}\left \{ \mathbb{E}_{F_{Y^n}}[Z]-D(F_{Y^n}\parallel F_{X^n})\right \} = \ln  \mathbb{E}_{F_{X^n}}\left [\exp\{Z\}\right].
\]
Continuing with the chain of equalities from (\ref{eqn:Renyi_Entropy}) we get
\begin{eqnarray}
\lefteqn {\sup_{F_{Y^n}}\bigg \{F_{Y^n}\rho H_{\frac{1}{1+\rho}}(P_{X^n}^{\prime}) + F_{Y^n}^{c}\rho nR - D(F_{Y^n}\parallel F_{X^n}) \bigg \}  }\nonumber \\
&=&\hspace*{-.1in}\ln \left\{ F_{X^n}^{c}\exp\{nR\rho\} + F_{X^n}\left (\sum_{i=1}^{M}{P_{X^n}^{\prime}}^{\frac{1}{1+\rho}}(i)\right )^{1+ \rho} \right\} \nonumber \\
\label{eqn:maximum-varational-difference}
&=&\ln \left \{ F_{X^n}^{c}\exp\{nR\rho\} + \left (\sum_{i=1}^{M}P_{X^n}^{\frac{1}{1+\rho}}(i)\right )^{1+ \rho} \right  \}.
\end{eqnarray}
Finally normalize both sides of (\ref{eqn:maximum-varational-difference}) by $n$, take limit infimum, and apply \cite[Lemma 1.2.15]{1998LDTA_DemZei}, which states that the exponential rate of a sum is governed by the maximum of the individual terms' exponential rates, to get the desired result.
\end{IEEEproof}

In the subsequent subsections we further lower bound each of the two terms under max on the right-hand side of (\ref{eqn:AymptoticLowerBound}). For an arbitrary source we first recall the source coding error exponent. We also identify the growth rate of sum of exponentiated probabilities of the correct decoding set. We then relate these to the terms in the lower bound obtained in (\ref{eqn:AymptoticLowerBound}). We largely follow the approach and notation of Iriyama \cite{200105TIT_Iri}, which we now describe.

Given $\mathbf{X}=\left(P_{X^n}: n \in \mathbb{N}\right)$ and $\mathbf{Y}=\left(P_{Y^n}: n \in \mathbb{N}\right)$,  we define the upper divergence $D_u(\cdot\parallel \cdot)$ and lower divergence $D_l(\cdot\parallel \cdot)$ by
\[
D_u(\mathbf{Y}\parallel \mathbf{X}):= \limsup_{n \rightarrow \infty}\frac{1}{n}D(P_{Y^n}\parallel P_{X^n})
\]
\[
D_l(\mathbf{Y}\parallel \mathbf{X}):= \liminf_{n \rightarrow \infty}\frac{1}{n}D(P_{Y^n}\parallel P_{X^n}).
\]
For a $\mathbf{Y}=\left(P_{Y^n}:n \in \mathbb{N}\right)$, denote the \textit{spectral sup-entropy-rate} \cite[Sec. II]{200009TIT_Han}, \cite{2003ISMIT_Han} as
\begin{equation*}
\overline {H}(\mathbf{Y}):=\inf \left \{\theta: \displaystyle \lim_{n \rightarrow \infty}\Pr\left\{\frac{1}{n}\ln \frac{1}{P_{Y^n}(Y^n)}> \theta\right \}=0\right\},
\end{equation*}
and the \textit{spectral inf-entropy-rate} as
\begin{equation*}
\underline {H}(\mathbf{Y}):=\sup \left \{\theta: \displaystyle \lim_{n \rightarrow \infty}\Pr\left\{\frac{1}{n}\ln \frac{1}{P_{Y^n}(Y^n)}< \theta\right
\}=0\right\}.
\end{equation*}
Also define, as in \cite[Sec. II]{200105TIT_Iri}, the following quantity which determines the performance under mismatched compression:
\begin{equation*}
\underline{R}(\mathbf{Y},\mathbf{X})\hspace*{-.02in}:=\sup \left \{\theta:\hspace*{-.08in} \lim_{n \rightarrow \infty}\Pr\left \{\frac{1}{n}\ln \frac{1}{P_{X^n}(Y^n)} < \theta \right \}\hspace*{-.04in}=\hspace*{-.04in}0 \right \}.
\end{equation*}

\subsubsection{Decoding Error Exponent}
\label{subsubsec:DecodingErrorExponent}
In this subsection we recall the decoding error exponent for fixed-rate encoding of an arbitrary source. We identify the first term in (\ref{eqn:AymptoticLowerBound}) as composed of the exponent of minimum probability of decoding error, and obtain a lower bound for it, or alternatively an  upper bound on the error exponent. This is made precise in the following definitions.

By an $(n,M_n,\epsilon_n)$-code we mean an encoding mapping
\[
\phi_n: \mathbb{X}^n \rightarrow \{1,2,\cdots,M_n\}
\]
and a decoding mapping
\[
\psi_n: \{1,2,\cdots M_n\} \rightarrow \mathbb{X}^n
\]
with probability of error $\epsilon_n:=\Pr\{\psi_n(\phi_n(X^n))\neq X^n\}$. $R$ is $r$-achievable if for all $\eta >0$ there exists a sequence of $(n,M_n,\epsilon_n)$-codes such that
\begin{eqnarray}
\label{eqn:error-liminf-exponent}
\limsup_{n \rightarrow \infty}\frac{1}{n}\ln \frac{1}{\epsilon_n} & \geq & r \\
\label{eqn:R-rate}
\limsup_{n \rightarrow \infty}\frac{1}{n}\ln M_n & \leq & R + \eta.
\end{eqnarray}
The {\em infimum fixed-length coding rate} for exponent $r$ is
     \begin{equation*}
     \label{def:inf-rate-erorr}
     \hat{R}(r|\mathbf{X})=\inf \{R: R~ \textnormal{is} ~r\textnormal{-achievable} \}.
     \end{equation*}
On the other hand, the {\em supremum fixed-length coding exponent} for rate $R$ is
    \begin{equation*}
   \label{def:sup-exponent-erorr}
     \hat{E}(R|\mathbf{X})=\sup \{r:  R~ \textnormal{is} ~r\textnormal{-achievable}  \}.
    \end{equation*}
See Iriyama \cite{200105TIT_Iri} and Han \cite[Sec. 1.9]{2003ISMIT_Han} for a pessimistic definition for fixed rate source coding, i.e., the liminf in place of limsup in (\ref{eqn:error-liminf-exponent}). See also Iriyama \& Ihara \cite{200110IEICE_IriIha} for both the pessimistic and optimistic definitions. These works obtained bounds on the infimum coding rate. In particular, Iriyama \cite[Eqn. (13)]{200105TIT_Iri},
Iriyama \& Ihara \cite[Eqn. (12)]{200110IEICE_IriIha}
obtained lower bounds on the infimum coding rate $\hat{R}(r|\mathbf{X})$ under the optimistic definition, the definition of interest to us.
We however work with the error exponent, and obtain an upper bound on supremum coding exponent. This suffices to lower bound the first term in (\ref{eqn:AymptoticLowerBound}).

Clearly, $M_n = \lfloor \exp \{ nR \}\rfloor$ satisfies (\ref{eqn:R-rate}), and with
$$r_0 = \limsup_{n \rightarrow \infty}  \frac{1}{n}\log \frac{1}{F_{X^n}^{c}},$$
$R$ is $r_0$-achievable. It follows from the definition of $\hat{E}(R|\mathbf{X})$ that
\[
\displaystyle \limsup_{n \rightarrow \infty}  \frac{1}{n}\ln \frac{1}{F_{X^n}^{c}} \leq \hat{E}(R|\mathbf{X})
\]
so that
\[
\displaystyle \liminf_{n \rightarrow \infty}  \frac{1}{n}\ln F_{X^n}^{c} \geq -\hat{E}(R|\mathbf{X}).
\]
The following proposition upper bounds the supremum coding exponent.
\begin{proposition}
\label{prop:sup-exponent-error-liminf}
For any rate $R>0$,
\begin{equation}
\label{eqn: EhatR-upperbound}
\hat{E}(R|\mathbf{X})\leq
\inf_{\mathbf{Y} :\underline{H}(\mathbf{Y}) > R} D_u(\mathbf{Y}\parallel \mathbf{X}).
\end{equation}
\end{proposition}

\begin{IEEEproof}
See Appendix \ref{subsec:sup-exponent-error-liminf-proof}.
\end{IEEEproof}
\begin{remark}
\label{remark:ErrorExponentEmptySet}
When $R \geq \ln |\mathbb{X}|$, the probability of decoding error $\epsilon_n = 0$,
so that $\hat{E}(R|\mathbf{X}) = +\infty$. The right-hand side is an infimum over an empty set and is $+\infty$ by convention, and the proposition holds for such $R$ as well.

One can also show the alternative bound
\begin{equation}
\label{eqn: EhatR-upperbound-alternative}
\hat{E}(R|\mathbf{X})\leq
\inf_{\mathbf{Y} :\underline{R}(\mathbf{Y},\mathbf{X})-D_u(\mathbf{Y}\parallel \mathbf{X})> R} D_u(\mathbf{Y}\parallel \mathbf{X}).
\end{equation}
See the end of Appendix \ref{subsec:sup-exponent-error-liminf-proof} on how to prove this. This result would be the functional inverse of Iriyama's \cite[Eqn. (13)]{200105TIT_Iri}, while Proposition \ref{prop:sup-exponent-error-liminf} is the functional inverse of Iriyama \& Ihara's \cite[Eqn. (12)]{200110IEICE_IriIha}. Proposition \ref{prop:sup-exponent-error-liminf}, as we will soon see, provides a more natural extension of Arikan \& Merhav's expression for $E(R,\rho)$ to general sources.
\hfill \IEEEQEDopen
\end{remark}

\subsubsection{Correct Decoding Exponent}
\label{subsubsec:CorrectDecodingExponent}
We now study a generalization of the exponential rate for probability of correct decoding.

For a given $(n,M_n,\epsilon_n)$-code, let
\[
A_n:=\{x^n \in \mathbb{X}^n:\psi_n(\phi_n(x^n))= x^n\}
\] denote the set of correctly decoded sequences.
For a given $\rho>0$, $R$ is $(r,\rho)$-admissible if for every $\eta >0$ there exists a sequence of $(n,M_n,\epsilon_n)$-codes such that
\begin{equation}
\label{eqn:correct-liminf-exponent}
(1+\rho) \liminf_{n\rightarrow \infty}\frac{1}{n}\ln \sum_{x^n \in A_n}P_{X^n}^{\frac{1}{1+\rho}}(x^n) \geq r
\end{equation}
\begin{equation}
\limsup_{n \rightarrow \infty}\frac{1}{n}\ln M_n \leq R + \eta.
\end{equation}
Unlike the exponent for the probability of error, here $r$ can be positive or negative.
The {\em infimum fixed-length admissible rate} for a given $r$ and $\rho > 0$ is
\begin{equation*}
\label{def:inf-rate-correct}
R^*(r,\rho|\mathbf{X})=\inf \{R: R~ \textnormal{is}~(r,\rho)\textnormal{-admissible} \}.
\end{equation*}
It is easy to see that the set $\{R: R \mbox{ is } (r,\rho)\textnormal{-admissible} \}$ is closed and so $R^*(r,\rho|\mathbf{X})$ is $(r,\rho)$-admissible.\\
The {\em supremum fixed-length coding exponent} for a given $R$ and $\rho$ is
\begin{equation*}
\label{def:sup-exponent-correct}
E^*(R,\rho|\mathbf{X})=\sup \{r: R~ \textnormal{is}~ (r,\rho)\textnormal{-admissible} \}.
\end{equation*}
\begin{remark}
The choice of limit infimum in (\ref{eqn:correct-liminf-exponent}) makes the definition of admissibility pessimistic. For $\rho\downarrow 0$, the above definitions reduce to the special case of exponential rate for probability of correct decoding (see \cite[Sec. 1.10]{2003ISMIT_Han}). \hfill \IEEEQEDopen
\end{remark}

Clearly, $A_n$ should be $T_R(X^n)$ to maximize the left-hand side of (\ref{eqn:correct-liminf-exponent}), and hence
\[
 E^*(R,\rho|\mathbf{X})=(1+ \rho)\liminf_{n \rightarrow \infty}\frac{1}{n}\ln \sum_{x^n \in T_R(X^n)}P_{X^n}^{\frac{1}{1+\rho}}(x^n).
\]
The following proposition gives an expression for $E^*(R,\rho|\mathbf{X})$ and generalizes \cite[Thm. 4]{200105TIT_Iri} to any arbitrary $\rho >0$. En route to its derivation we find the expression for $R^*(r,\rho|\mathbf{X})$.
\begin{proposition}
\label{prop:inf-rate-correct}
For any $\rho > 0$, we have
\begin{equation}
\label{eqn:correct-rate-expression-liminf}
R^{*}(r,\rho|\mathbf{X})=\inf_{\mathbf{Y}:E_l(\mathbf{Y},\mathbf{X},\rho)\geq r} \overline{H}(\mathbf{Y})
\end{equation}
\begin{equation}
\label{eqn:correct-exponent-expression-liminf}
E^*(R,\rho|\mathbf{X})=\sup_{\mathbf{Y}: \overline{H}(\mathbf{Y})\leq R}{E}_l(\mathbf{Y},\mathbf{X},\rho).
\end{equation}
\end{proposition}

\begin{IEEEproof}
See Appendix \ref{subsec:inf-rate-correct-proof}.
\end{IEEEproof}

\subsection{Summary of Bounds on $E_u^s$ and $E_l^s$}
\label{subsec:UpperLowerBound}
We now combine Propositions \ref{prop:UpperBound}-\ref{prop:inf-rate-correct} of the previous subsections to obtain the main result of the paper.
\begin{theorem}
For a given $\rho > 0$ and $R>0$,
\begin{eqnarray}
\label{eqn:FinalBounds}
\lefteqn { \max \bigg \{\rho R-\inf_{\mathbf{Y} :\underline{H}(\mathbf{Y}) > R} D_u(\mathbf{Y}\parallel \mathbf{X}), } \nonumber\\
&& \quad \quad \quad \quad \quad \quad \sup_{\mathbf{Y}: \overline{H}(\mathbf{Y})\leq R}{E}_l(\mathbf{Y},\mathbf{X},\rho) \bigg \}  \nonumber\\
&\leq &E_l^s(R, \rho)\leq E_u^s(R, \rho)\nonumber\\
&\leq & \min_{ 0\leq \theta\leq \rho}    \left \{(\rho -\theta)R + \max_{\mathbf{Y}}E_u(\mathbf{Y},\mathbf{X},\theta)\right \}.
\end{eqnarray}
\end{theorem}

\begin{IEEEproof}
The last inequality was proved in Proposition \ref{prop:UpperBound}. Proposition \ref{prop:LowerBound} indicates that
\begin{eqnarray}
\lefteqn {E_l^s(R,\rho)  }  \nonumber\\
\label{eqn:liminf-log-sum}
&\geq&\max \bigg \{\rho R+\liminf_{n \rightarrow \infty}\frac{1}{n}\ln  F_{X^n}^{c} , \nonumber\\
&&\hspace{.5in}(1+ \rho)\liminf_{n \rightarrow \infty}\frac{1}{n}\ln \sum_{x^n \in T_R(X^n)}P_{X^n}^{\frac{1}{1+\rho}}(x^n) \bigg \} \nonumber\\
\label{eqn:substitute-definition}
&\geq&\max \left \{\rho R-\hat{E}(R|\mathbf{X}), E^*(R,\rho|\mathbf{X}) \right \} \\
\label{eqn:substitute-expression}
&\geq&\max \bigg \{\rho R-\inf_{\mathbf{Y} :\underline{H}(\mathbf{Y}) > R} D_u(\mathbf{Y}\parallel \mathbf{X}), \nonumber\\
&&\hspace*{1.3in}\sup_{\mathbf{Y}: \overline{H}(\mathbf{Y})\leq R}{E}_l(\mathbf{Y},\mathbf{X},\rho) \bigg \},
\end{eqnarray}
where (\ref{eqn:substitute-definition}) follows from the lower bound on $\hat{E}(R|\mathbf{X})$ and the definition of $E^*(R,\rho|\mathbf{X})$, and (\ref{eqn:substitute-expression}) from Propositions \ref{prop:sup-exponent-error-liminf} and \ref{prop:inf-rate-correct}.
\end{IEEEproof}

\section{Examples}
\label{sec:Examples}
In this section we evaluate the bounds for some examples where they are tight, and recover some known results.
\begin{example}[Perfect Secrecy]
First consider the perfect secrecy case, for example, $R\geq \ln |\mathbb{X}|$. Because of Remark \ref{remark:ErrorExponentEmptySet} and because we may take $\theta=\rho$ in the upper bound in (\ref{eqn:FinalBounds}), the limiting exponential rate of guessing moments simplifies to
\begin{eqnarray*}
\sup_{\mathbf{Y}}{E}_l(\mathbf{Y},\mathbf{X},\rho)&\leq& E_l^s(R, \rho)\\
&\leq& E_u^s(R, \rho)\leq \max_{\mathbf{Y}}E_u(\mathbf{Y},\mathbf{X},\rho).
\end{eqnarray*}
On account of (\ref{eqn:liminf-El}) in Proposition \ref{prop: limit-variational-formula}, sup in the left-most term is achieved.
From Proposition \ref{prop: limit-variational-formula}, upper and lower bounds are $\rho$ times the liminf and limsup R\'{e}nyi entropy rates of order $\frac{1}{1+\rho}$. In a related work we proved in \cite[Prop. 7]{200812TRPME_HanSun} that whenever the {\em information spectrum} of the source satisfies the large deviation property with rate function $I$, the R\'{e}nyi entropy rate converges and limiting guessing exponent equals the Legendre-Fenchel dual of the scaled rate function $I_1(t):=(1+\rho)I(t)$, i.e.,
\[E_u^s(R, \rho)=E_l^s(R, \rho)=\sup_{ t \in \mathbb{R}}\{\rho t-I_1(t)\}.\]
In the next examples, we consider the case $R <\ln |\mathbb{X}|$.
\end{example}

\begin{example}[An iid source]
This example was first studied by Merhav \& Arikan \cite{199909TIT_MerAri}. Recall that an iid source is one for which $P_n(x^n)=\prod_{i=1}^{n}P_1(x_i)$, where $P_1$ denotes the marginal of $X_1$. We will now evaluate  each term in (\ref{eqn:FinalBounds}).

We first argue that
\begin{eqnarray}
\label{eqn:IIDErrorExpo}
\inf_{\mathbf{Y} :\underline{H}(\mathbf{Y}) > R} D_u(\mathbf{Y}\parallel \mathbf{X}) = \inf_{P_Y:H(P_Y)>R}D(P_Y \parallel P_1).
\end{eqnarray}
To prove that the left-hand side in (\ref{eqn:IIDErrorExpo}) is less than or equal to the right-hand side, let $P_{Y} \in \mathcal{M}(\mathbb{X})$ be such that $H(P_Y)>R$. Construct an iid source $\hat{\mathbf{Y}} = (P_{\hat{Y}^n}:n \in \mathbb{N})$ such that $P_{\hat{Y}_i}=P_Y$ for all $1\leq i \leq n$. The iid property easily implies that
\[D_u(\hat{\mathbf{Y}}\parallel\mathbf{X})=D(P_Y\parallel P_1),\]
and the law of large numbers for iid random variables yields
\begin{equation}
\label{eqn:IIDCOnverseRdiffD}
\underline{H}(\hat{\mathbf{Y}}) = H(P_Y) > R.
\end{equation}
From (\ref{eqn:IIDCOnverseRdiffD}), we have that the infimum on the left-hand side of (\ref{eqn:IIDErrorExpo}) is over a larger set. We can therefore conclude that ``$\leq$'' holds in (\ref{eqn:IIDErrorExpo}).

To prove ``$\geq$'' in (\ref{eqn:IIDErrorExpo}) we use the result (see \cite[Th. 1.7.2]{2003ISMIT_Han})
\[
  \underline{H}(\mathbf{Y}) \leq H_l(\mathbf{Y}) := \liminf_{n \rightarrow \infty}\frac{1}{n}H(P_{Y^n})
\]
to get that the infimum over a larger set is smaller, i.e.,
\begin{equation}
\label{eqn:AnotherExponentBound}
\inf_{\mathbf{Y} :\underline{H}(\mathbf{Y}) > R} D_u(\mathbf{Y}\parallel \mathbf{X})
\geq  \inf_{\mathbf{Y}:H_{l}(\mathbf{Y})>R}D_u(\mathbf{Y}\parallel \mathbf{X}).
\end{equation}
Because of (\ref{eqn:AnotherExponentBound}) it is sufficient to prove
\begin{equation}
\label{eqn:IIDAnotherBound}
\inf_{\mathbf{Y}:H_{l}(\mathbf{Y})>R}D_u(\mathbf{Y}\parallel \mathbf{X})\geq \inf_{P_Y:H(P_Y)>R}D(P_Y \parallel P_1).
\end{equation}
Let $\mathbf{Y}$ be such that $H_{l}(\mathbf{Y})>R$. Construct a source $\hat{\mathbf{Y}}$ such that, $P_{\hat{Y}_i}=P_{Y_i}$ for $1\leq i \leq n$ and $\hat{Y}_1,\hat{Y}_2,\cdots,\hat{Y}_n$ are independent. Let
$\mathbf{Z}$ be another source such that $Z_1,Z_2,\cdots ,Z_n$ is an iid sequence with distribution
\[P_{Z_j}=\frac{1}{n}\sum_{i=1}^{n}P_{Y_i},~~~j=1,2,\cdots,n.\]
As the marginals of $Y^n$ and $\hat{Y}^n$ with independent components are the same, it easily follows from the formula for Kullback-Leibler divergence that
\begin{eqnarray}
D(P_{Y^n}\parallel P_{X^n}) &=& D(P_{Y^n}\parallel P_{\hat{Y}^n})+ D(P_{\hat{Y}^n}\parallel P_{X^n}) \nonumber \\
&\geq& D(P_{\hat{Y}^n}\parallel P_{X^n}) \nonumber \\
& = & \sum_{i=1}^n  D(P_{\hat{Y}_i}\parallel P_1)  \nonumber \\
\label{eqn:DivergenceConvexBound}
&\geq&nD(P_{Z_1}\parallel P_1),
\end{eqnarray}
where (\ref{eqn:DivergenceConvexBound}) follows from the convexity of divergence. From the concavity of Shannon entropy, we also have
\begin{equation}
\label{eqn:EntropyConcaveBound}
\displaystyle H(P_{Y^n})\leq\sum_{i=1}^{n}H(P_{Y_i})\leq nH(P_{Z_1}).
\end{equation}
Normalize by $n$ take limsup in (\ref{eqn:DivergenceConvexBound}) and liminf in (\ref{eqn:EntropyConcaveBound}) to get $D_u(\mathbf{Y}\parallel \mathbf{X})\geq D(P_{Z_1}\parallel P_1)$ and $H(P_{Z_1})>R$ for a $P_{Z_1}$ that is a limit point of the sequence $(n^{-1}\sum_{i=1}^{n}P_{Y_i},n \in \mathbb{N})$. From these we conclude that (\ref{eqn:IIDAnotherBound}) holds. This proves (\ref{eqn:IIDErrorExpo}).

Following a similar procedure as above, we can bound the other terms in (\ref{eqn:FinalBounds}) for an iid source as
\begin{eqnarray}
\label{eqn:IIDElLBound}
\lefteqn {\sup_{\mathbf{Y}:\overline{H}(\mathbf{Y})\leq R}E_{l}(\mathbf{Y},\mathbf{X},\rho)}\nonumber \\
&&\geq \sup_{P_{Y}:H(P_Y)\leq R}\{\rho H(P_Y)-D(P_Y\parallel P_1)\}
\end{eqnarray}
and
\begin{eqnarray}
\label{eqn:IIDEuUBound}
\sup_{\mathbf{Y}}E_{u}(\mathbf{Y},\mathbf{X},\theta)= \sup_{P_{Y}}\{\theta H(P_Y)-D(P_Y\parallel P_1)\}.
\end{eqnarray}
Substitution of (\ref{eqn:IIDErrorExpo}) and (\ref{eqn:IIDElLBound}) in the lower bound of (\ref{eqn:FinalBounds}) yields
\begin{eqnarray}
\lefteqn { E_l^s(R,\rho) \geq \max \bigg\{\rho R- \inf_{P_Y:H(P_Y)>R}D(P_Y \parallel P_1),}\nonumber \\
&&\hspace*{.75in}\sup_{P_{Y}:H(P_Y)\leq R}\{\rho H(P_Y)-D(P_Y\parallel P_1)\}\bigg\} \nonumber\\
&&\nonumber\\
\label{eqn:IIDGuessExponentLBound}
&&\hspace{.255in}=\sup_{P_Y}\left\{\rho \min\{H(P_Y),R\}-D(P_Y\parallel P_1)\right\}.
\end{eqnarray}
Similarly substitution of  (\ref{eqn:IIDEuUBound}) in the upper bound of (\ref{eqn:FinalBounds}) yields
\begin{eqnarray}
\lefteqn {E_u^s(R,\rho)} \nonumber\\
&&\leq \min_{ 0\leq \theta\leq \rho}    \left \{(\rho -\theta)R + \sup_{P_{Y}}\{\theta H(P_Y)-D(P_Y\parallel P_1)\} \right \} \nonumber  \\
&& =   \sup_{P_{Y}} \bigg \{\min_{ 0\leq \theta\leq \rho} \{ (\rho -\theta)R + \theta H(P_Y)\} -D(P_Y\parallel P_1) \bigg \} \nonumber\\
\label{eqn:MinSupInterchange}
&&\\
\label{eqn:IIDGuessExponentUBound}
&&= \sup_{P_{Y}} \left \{\rho \min\{H(P_Y),R\} -D(P_Y\parallel P_1) \right \},
\end{eqnarray}
where the interchange of sup and min in (\ref{eqn:MinSupInterchange}) holds because the function within braces is linear in $\theta$ and concave in $P_Y$. From (\ref{eqn:IIDGuessExponentLBound}) and (\ref{eqn:IIDGuessExponentUBound}), we recover Merhav \& Arikan's result (\ref{eqn:IIDGuessExponent}) for an iid source \cite[Eqn. (3)]{199909TIT_MerAri}.
\end{example}
\begin{example}[Markov source]
\label{example:Markov}
In this example we focus on an irreducible stationary Markov source taking values on $\mathbb{X}$ and having a transition probability matrix $\pi$.

Let $\mathcal{M}_s(\mathbb{X}^2)$ denote the set of {\em stationary} PMFs defined by
\begin{eqnarray*}
  \lefteqn{ \mathcal{M}_s \left(\mathbb{X}^2\right) = \Big\{ Q \in \mathcal{M} \left(\mathbb{X}^2\right) : } \\
  && ~~~~~~~~~~~~\sum_{x_1 \in \mathbb{X}} Q(x_1, x) = \sum_{x_2 \in \mathbb{X}} Q(x,x_2), \forall x \in \mathbb{X} \Big\}.
\end{eqnarray*}
Denote the common marginal by $q$ and let
\[
  \eta(\cdot \mid x_1) := \left\{
      \begin{array}{cl}
        Q(x_1, \cdot) / q(x_1), &  \mbox{if } q(x_1) \neq 0, \\
        1/|\mathbb{X}|, & \mbox{otherwise}.
      \end{array}
  \right.
\]
We may then denote $Q = q \times \eta$, where $q$ is the distribution of $X_1$ and $\eta$ the conditional distribution of $X_2$ given $X_1$.
Following steps similar to the iid case, we have

\[E_u^s=E_l^s = \sup_{Q \in \mathcal{M}_s(\mathbb{X}^2)} \Big\{\rho \min\{H(\eta \mid q),R\} - D(\eta \parallel \pi \mid q)\Big\},\]
where
\[
  H(\eta \mid q) := \sum_{x \in \mathbb{X}} q(x) H(\eta(\cdot \mid x)).
\]
is the conditional one-step entropy, and
\[
 D(\eta \parallel \pi \mid q) = \sum_{x_1 \in \mathbb{X}} q(x_1) D(\eta(\cdot \mid x_1) \parallel \pi(\cdot \mid x_1)).
\]
For a unifilar source the underlying state space forms a Markov chain and the entropy and divergence of the source equals those of the underlying  Markov state space source \cite[Thm. 6.4.2]{1965xxIT_Ash}. The arguments for the Markov source are now directly applicable to a unifilar source.
\end{example}

\section{Conclusion}
We saw the close connection between the problem of guessing a source realization given a cryptogram and the problem of compression with saturated exponential costs. The latter is a modification of a problem posed by Campbell \cite{Campbell-1}. Moreover, the exponents for both these problems coincide. This exponent is determined by the error exponent and a generalization of correct decoding exponent for fixed length block source codes.

We end this paper with some open questions.
\begin{itemize}
\item The equivalence between guessing and compression exploits the finite alphabet size assumption. Can this be relaxed?
\item How do the results of this paper extend to the case with receiver side information? Can the result of Hayashi \& Yamamoto be extended to general sources?
\item If guessing to within a distortion is allowed, can the result of Merhav \& Arikan \cite{Arikan-Merhav} be extended to general sources? Both cases of perfect secrecy and key-rate constrained secrecy remain open.
\end{itemize}

\label{sec:Conclusion}

\appendices

\label{sec:Appendix}
\section{Proof of Proposition \ref{prop:key:OptimalEncryption}}
\label{subsec:ProofOptimalEncryption}
Let $P_n$ be any PMF on $\mathbb{X}^n$.
Enumerate the elements of $\mathbb{X}^n$ from 1 to $|\mathbb{X}|^n$ in the decreasing order of their $P_n$-probabilities. Let $M = \exp\{nR\}$ denote the number of distinct key strings. For convenience, we shall assume that $M$ is a power of 2 so that the number of key bits $k = nR/(\ln 2)$ is an integer. The general case will be easily handled towards the end of this section.

If $M$ does not divide $|\mathbb{X}|^n$, append a few dummy messages of zero probability to make the number of messages $N$ a multiple of $M$. Further, index the messages from 0 to $N-1$. Henceforth, we identify a message $x^n$ by its index.

Divide the messages into groups of $M$ so that message $m$ belongs to group $T_j$, where $j = \lfloor m/M \rfloor$, and $\lfloor \cdot \rfloor$ is the floor function. Enumerate the key streams from 0 to $M-1$, so that $0 \leq u \leq M-1$. The function $f_n$ is now defined as follows.
For $m = jM + i$ set
\[
  f_n(jM+i, u) \stackrel{\Delta}{=} jM + \left( i \oplus u \right),
\]
where $i \oplus u$ is the bit-wise XOR operation. Thus messages in group $T_j$ are encrypted to messages in the same group. The index
$i$ identifying the specific message in group $T_j$, i.e., the last $k = nR/(\ln 2)$ bits of $m$, are encrypted via bit-wise XOR with the key
stream. Given $u$ and the cryptogram, decryption is clear -- perform bit-wise XOR with $u$ on the last $nR/(\ln 2)$ bits of $y$.

Given a cryptogram $y$, the only information that the attacker gleans is that the message belongs to the group determined by $y$.
Indeed, if $y \in T_j$, then
\[
  P_n \left\{ Y = y \right\} = \frac{1}{M} P_n \left\{ X^n \in T_j \right \},
\]
and therefore
\[
  P_n \left\{ X^n = m \mid Y=y \right\} = \left\{
    \begin{array}{ll}
     \frac{ P_n \left\{ X^n = m \right \} }{ P_n \left\{ X^n \in T_j \right\}
     }, & \lfloor m/M \rfloor = j, \\
     0, & \mbox{otherwise},
    \end{array}
  \right.
\]
which decreases with $m$ for $m \in T_j$, because of our enumeration in the decreasing order of probabilities, and is 0 for $m \notin T_j$. The attacker's best strategy $G_{f_n}(\cdot \mid y)$ is therefore to
restrict his guesses to $T_j$ and guess in the order $jM, jM+1, \cdots, jM + M-1$. Thus, when $x^n=jM+i$, the optimal attack
strategy requires $i+1$ guesses.

We now analyze the performance of this attack strategy as follows.
\begin{eqnarray}
  \nonumber
  \lefteqn{ \mathbb{E} \left[ G_{f_n}(X^n|Y)^{\rho} \right] } \\
    \nonumber
    &  =   & \sum_{j=0}^{N/M-1} \sum_{i=0}^{M-1} P_n \{ X^n = j M + i \} (i+1)^{\rho} \\
    \label{eqn:sumProbM}
    & \geq & \sum_{j=0}^{N/M-1} \sum_{i=0}^{M-1} P_n \{ X^n = (j+1)M - 1 \} (i+1)^{\rho} \\
    \label{eqn:sumIntBound}
    & \geq & \sum_{j=0}^{N/M-1} P_n \{ X^n = (j+1)M - 1 \} \frac{ M^{1+ \rho}} {1+\rho} \\
    \nonumber
    & \geq & \frac{1}{1+\rho} \sum_{j=0}^{N/M-1} \sum_{i=0}^{M-1} P_n \{ X^n = (j+1)M + i \} M^{\rho} \\
    \label{eqn:sumProbMBound}
        & & \\
    \label{eqn:G*lowerBound}
    & =    & \frac{1}{1+\rho} \sum_{m=M}^{N-1} P_n \{ X^n = m \} M^{\rho}
\end{eqnarray}
where (\ref{eqn:sumProbM}) follows because the arrangement in the decreasing order of probabilities implies that
\[
  P_n \{ X^n = jM + i \} \geq P_n \{ X^n = (j+1)M - 1 \}
\]
for $i = 0, \cdots, M-1$. Inequality (\ref{eqn:sumIntBound}) follows
because
\[
  \sum_{i=0}^{M-1} (i+1)^{\rho} = \sum_{i=1}^{M} i^{\rho} \geq
  \int_0^M z^{\rho}~dz = \frac{M^{1+\rho}}{1+\rho}.
\]
Inequality (\ref{eqn:sumProbMBound}) follows because the decreasing
probability arrangement implies
\[
  P_n \{X^n = (j+1)M - 1 \} \geq \frac{1}{M} \sum_{i=0}^{M-1} P_n \{ X^n = (j+1)M + i\}.
\]
Inequality (\ref{eqn:G*lowerBound}) follows because we take $P_n(X^n = m) = 0$ for all the further dummy messages with indices $m > N$. Thus (\ref{eqn:G*lowerBound}) implies that
\begin{eqnarray}
  \nonumber
  \lefteqn{ \sum_{m=0}^{N-1} P_n \{ X^n = m \} \left( \min \{ m+1, M \}
  \right)^{\rho} } \\
  \nonumber
  & = & \hspace*{-.05in} \sum_{m=0}^{M-1} P_n \{ X^n = m \} (m+1)^{\rho} + \sum_{m=M}^{N-1} P_n \{ X^n = m \}
  M^{\rho} \\
  \nonumber
  & \leq & \hspace*{-.05in} \mathbb{E} \left[ G_{f_n}(X^n|Y)^{\rho} \right] + (1 + \rho) \mathbb{E} \left[ G_{f_n}(X^n|Y)^{\rho}
  \right] \\
  \label{eqn:expectedG*Bound}
  & = & \hspace*{-.05in} (2 + \rho) \mathbb{E} \left[ G_{f_n}(X^n|Y)^{\rho} \right].
\end{eqnarray}
Let $G$ be the guessing function that guesses in the decreasing order of $P_n$-probabilities without regard to $Y$, i.e., $G(m) =
m+1$. Let $L_{G}$ be the associated length function, given in Lemma \ref{lema:GuessLengthRelation}. Now use (\ref{eqn:expectedG*Bound}) and Lemma \ref{lema:GuessLengthRelation} to
get
\begin{eqnarray}
  \lefteqn{ \mathbb{E} \left[ G_{f_n}(X^n|Y)^{\rho} \right] } \nonumber \\
  & \geq &
  \frac{1}{2+\rho} \mathbb{E} \left[ \left( \min \left\{ G(X^n), M \right\} \right)^{\rho}
  \right] \nonumber \\
  & \geq &
  \frac{1}{2+\rho} \mathbb{E} \left[ \left( \min \left\{ \frac{\exp_2\{L_{G}(X^n)\}}{2 c_n}, M \right\} \right)^{\rho}
  \right] \nonumber \\
  & \geq &
  \frac{1}{(2 c_n)^{\rho} (2+\rho)} \mathbb{E} \left[ \exp\left\{\rho \min \left\{ L_{G}(X^n) \ln 2, nR \right\} \right\} \right], \nonumber \\
  \label{eqn: EG-LB}
  & &
\end{eqnarray}
where the last inequality follows by pulling out $2 c_n$ and recognizing that $2 c_n M \geq M \geq \exp\{nR\}$.
Since $G_{f_n}$ is the strategy that minimizes $\mathbb{E} \left[G(X^n \mid Y)^{\rho} \right]$ , the proof is complete for the cases when $k = nR/(\ln 2)$ is an integer.

When $nR/(\ln 2)$ is not an integer, choose $k = \lceil nR/(\ln 2) \rceil$. Then $M = \exp_2\{k\} \geq \exp\{ nR \}$, and it immediately follows that inequality (\ref{eqn: EG-LB}) continues to hold. This completes the proof. \hfill \IEEEQEDclosed

\section{Proof of Proposition \ref{prop: limit-variational-formula}}
\label{subsec:limit-variational-formula-proof}
We begin with the following lemma. Recall that $\mathcal{M}(\mathbb{X})$ is the set of all probability measures on $\mathbb{X}$ and $\mathcal{M}(B)$ the subset of $\mathcal{M}(\mathbb{X})$ with support set $B \subseteq \mathbb{X}$:\[\mathcal{M}(B)=\{\nu \in \mathcal{M}(\mathbb{X}) : \nu(B)=1\}.\]
\begin{lemma}
\label{lema: variational-lemma}
For any $\rho > 0,\mu \in \mathcal{M}(\mathbb{X})$ and $B \subseteq \mathbb{X}$
\[(1+ \rho)\ln \sum_{x \in B} \mu^{\frac{1}{1+\rho}}(x)= \max_{\nu \in M(B)}\{\rho H(\nu)-D(\nu\parallel\mu)\}.\]
\end{lemma}

\begin{remark}
\cite[Lemma 1]{200105TIT_Iri} is the special case when $\rho=0$. \hfill \IEEEQEDopen
\end{remark}

\begin{IEEEproof}
Let $\mu_B(x)=\frac{\mu(x)}{\mu(B)}1\{x \in B\}$. We then have
\begin{eqnarray}
\lefteqn {(1+ \rho)\ln \sum_{x \in B} \mu^{\frac{1}{1+\rho}}(x)} \nonumber\\
\label{eqn:normalisedmeasure}
&= &(1+ \rho)\ln \sum_{x \in B} {\mu_B}^{\frac{1}{1+\rho}}(x) + \ln \mu(B) \nonumber\\
&=&(1+\rho)\max_{\nu \in  \mathcal{M}(B)}\bigg\{ \lefteqn{ \sum_{x\in B} \frac{\rho}{1+\rho}\nu(x)\ln \frac{1}{\mu_B (x)} } \nonumber\\
\label{eqn:variational-inequality}
&& \hspace{1in} - D(\nu\parallel\mu_B) \bigg\} +\ln \mu(B) \\
&=&(1+\rho)\max_{\nu \in \mathcal{M}(B)}\bigg\{ \lefteqn{ \frac{\rho}{1+\rho}\left\{H(\nu)+D(\nu\parallel\mu)\right \}} \nonumber\\
&& \hspace{1in}  -D(\nu\parallel \mu)\bigg\} \\
\label{eqn:Renyi-formaula}
&=& \max_{\nu \in \mathcal{M}(B)}\left\{\rho H(\nu)-D(\nu\parallel\mu)\right\}.
\end{eqnarray}
where (\ref{eqn:variational-inequality}) follows from the variational formula for R\'{e}nyi entropy of $\mu_B$. The maximum achieving distribution in (\ref{eqn:Renyi-formaula}) is $\mu^* \in \mathcal{M}(B)$ given by \[\mu^*(x)=\frac{\mu^{\frac{1}{1+\rho}}(x)}{\sum_{y \in B}\mu^{\frac{1}{1+\rho}}(y)}1\{x \in B\},\]
a fact that is easily verified via direct substitution.
\end{IEEEproof}

We now prove (\ref{eqn:liminf-El}); proof of (\ref{eqn:limsup-Eu}) is similar and therefore omitted. We begin by showing ``$\leq$'' in (\ref{eqn:liminf-El}). Let $\mathbf{X}^*=(P_{X^n}^*:  n \in \mathbb{N}) \in  \mathcal{M}(\mathbf{B})$ be as defined in (\ref{eqn:macachiev-sequnce-distribution}). It is straightforward to verify by direct substitution that
\[(1+ \rho)\ln \sum_{x^n \in B_n} P_{X^n}^{\frac{1}{1+\rho}}(x^n)=\rho H(P_{X^n}^*)-D(P_{X^n}^*\parallel P_{X^n}).\]
Normalize by $n$ and take limit infimum, and use the definition of $E_l(\mathbf{X}^*,\mathbf{X},\rho)$ to get
\begin{eqnarray}
\lefteqn {(1+\rho)\liminf_{n \rightarrow \infty}\frac{1}{n}\ln \displaystyle \sum_{x^n \in B_n}P_{X^n}^{\frac{1}{1+\rho}}(x^n)}\nonumber\\
\label{eqn:PoorLiminfElLowerBound}
&=& E_l(\mathbf{X}^*,\mathbf{X},\rho) \\
&\leq& \max_{\mathbf{Y} \in \mathcal{M}(\mathbf{B})}E_l(\mathbf{Y},\mathbf{X},\rho). \nonumber
\end{eqnarray}

To prove ``$\geq$'' in (\ref{eqn:liminf-El}), let $\mathbf{Y}=(P_{Y^n}:n \in \mathbb{N}) \in  \mathcal{M}(\mathbf{B}) $ be an arbitrary sequence. We may assume that for all sufficiently large $n$, $P_{Y^n}\ll P_{X^n}$ holds; otherwise $E_l(\mathbf{Y},\mathbf{X},\rho)=-\infty$ and the inequality ``$\geq$" holds automatically. Define $\mathbf{Y}^*=(P_{Y^n}^*:n \in \mathbb{N}) \in  \mathcal{M}(\mathbf{B})$ by
\[P_{Y^n}^*(y^n)=\frac{P_{Y^n}(y^n)}{P_{Y^n}(B_n)}1\{y^n \in B_n\}.\]
It is clear that $P_{Y^n}^* \in \mathcal{M}(B_n)$ for every $n$. From Lemma \ref{lema: variational-lemma}, we have
\begin{eqnarray}
\lefteqn { (1+\rho)\ln \displaystyle \sum_{x^n \in B_n}P_{X^n}^{\frac{1}{1+\rho}}(x^n) }\nonumber \\
&=&\max_{P_{Y^n} \in \mathcal{M}(B_n)}\left\{\rho H(P_{Y^n})-D(P_{Y^n}\parallel P_{X^n})\right\} \nonumber \\
\label{eqn:renyi-expo-lowerbound}
&\geq& \rho H(P_{Y^n}^*)-D(P_{Y^n}^*\parallel P_{X^n}).
\end{eqnarray}
We now study each term on the right-hand side of (\ref{eqn:renyi-expo-lowerbound}). The entropy term is lower bounded as follows:
\begin{eqnarray}
\lefteqn {\rho H(P_{Y^n}^*)}\nonumber\\
&=& \frac{\rho}{{P_{Y^n}(B_n)}} \left \{\sum_{x^n \in B_n} P_{Y^n}(x^n)\ln \frac{1}{P_{Y^n}(x^n)}\right \} \nonumber\\
&& \hspace{1.5in} + ~ \rho \ln P_{Y^n}(B_n)\nonumber\\
&=& \frac{\rho}{{P_{Y^n}(B_n)}} \left \{H(P_{Y^n})- \hspace*{-.05in} \sum_{x^n \in B_n^c} P_{Y^n}(x^n)\ln \frac{1}{P_{Y^n}(x^n)} \right\} \nonumber\\
&&\hspace{1.5in} + ~ \rho \ln P_{Y^n}(B_n) \nonumber\\
&=& \frac{\rho}{{P_{Y^n}(B_n)}}  \bigg \{H(P_{Y^n})-P_{Y^n}(B_n^c)H(P_{Y^n}|B_n^c) \nonumber\\
&&\hspace{.6in} + ~P _{Y^n}(B_n^c)\ln P_{Y^n}(B_n^c) \bigg \} + \rho \ln P_{Y^n}(B_n) \nonumber\\
&\geq& \frac{\rho}{{P_{Y^n}(B_n)}} \bigg \{H(P_{Y^n})-P_{Y^n}(B_n^c)n\ln |\mathbb{X}| \nonumber\\
&&\hspace{.6in} + ~ P_{Y^n}(B_n^c)\ln P_{Y^n}(B_n^c) \bigg \}+\rho \ln P_{Y^n}(B_n). \nonumber \\
\label{eqn:bound-entropy}
&&
\end{eqnarray}
The divergence term is upper bounded, as in the proof of Iriyama's \cite[Prop. 1]{200105TIT_Iri}, as follows:
\begin{eqnarray}
\lefteqn {D(P_{Y^n}^*\parallel P_{X^n})}\nonumber\\
&=& -\ln P_{Y^n}(B_n) \nonumber \\
& & \hspace*{.5in} + ~ \frac{1}{{P_{Y^n}(B_n)}} \sum_{x^n \in B_n}P_{Y^n}(x^n)\ln \frac{ P_{Y^n}(x^n)}{P_{X^n}(x^n)} \nonumber \\
&=& \lefteqn {-\ln P_{Y^n}(B_n)+ \frac{1}{{P_{Y^n}(B_n)}}D(P_{Y^n} \parallel P_{X^n})}\nonumber\\
&& \hspace{.5in} - ~ \frac{1}{{P_{Y^n}(B_n)}}\sum_{x^n \in B_n^c}P_{Y^n}(x^n)\ln \frac{ P_{Y^n}(x^n)}{P_{X^n}(x^n)}\nonumber \\
&\leq& \lefteqn { -\ln P_{Y^n}(B_n)+ \frac{1}{{P_{Y^n}(B_n)}}D(P_{Y^n} \parallel P_{X^n})} \nonumber\\
\label{eqn: log-bound}
&& \hspace{.5in} - ~ \frac{P_{Y^n}(B_n^c)-P_{X^n}(B_n^c)}{P_{Y^n}(B_n)} \\
\label{eqn: set-bound}
&\leq& \lefteqn {-\ln P_{Y^n}(B_n)+ \frac{1}{{P_{Y^n}(B_n)}}D(P_{Y^n} \parallel P_{X^n}) }\nonumber \\
&& \hspace{.5in} + ~ \frac{1}{P_{Y^n}(B_n)}.
\end{eqnarray}
To get (\ref{eqn: log-bound}), we used the fact that $\ln x \geq 1-\frac{1}{x}$ for all $x >0$ and in inequality (\ref{eqn: set-bound}) we used the relation
$$P_{Y^n}(B_n^c)-P_{X^n}(B_n^c) \geq -1.$$
Substitution of (\ref{eqn:bound-entropy}) and (\ref{eqn: set-bound}) in (\ref{eqn:renyi-expo-lowerbound}) and the fact that $\lim _{n \rightarrow \infty}P_{Y^n}(B_n)=1$ yield
\begin{eqnarray}
\lefteqn {(1+\rho)\liminf_{n \rightarrow \infty}\frac{1}{n} \ln \displaystyle \sum_{x^n \in B_n}P_{X^n}^{\frac{1}{1+\rho}}(x^n)} \nonumber \\
&\geq& \liminf_{n \rightarrow \infty}\frac{1}{n}\left \{ \rho H(P_{Y^n})-D(P_{Y^n} \parallel P_{X^n})- O(1)\right \} \nonumber\\
&=& E_l(\mathbf{Y},\mathbf{X},\rho). \nonumber
\end{eqnarray}
Since the choice of $\mathbf{Y}=(P_{Y^n}: n \in \mathbb{N}) \in \mathcal{M}(\mathbf{B})$ was arbitrary, we have proved ``$\geq$" in (\ref{eqn:liminf-El}).

From (\ref{eqn:PoorLiminfElLowerBound}) and (\ref{eqn:liminf-El}), the maximum is attained by $\mathbf{X}^*$, the distribution defined in (\ref{eqn:macachiev-sequnce-distribution}). This completes the proof.
\hfill \IEEEQEDclosed

\section{Proof of Proposition \ref{prop:sup-exponent-error-liminf}}
\label{subsec:sup-exponent-error-liminf-proof}
Iriyama \& Ihara showed the following lower bound on the infimum coding rate (\cite[Th.3, Eqn. (12)]{200110IEICE_IriIha}):
\begin{equation}
   \label{eqn: Rhatr-lowerbound}
   \sup_{\mathbf{Y} : D_u(\mathbf{Y} \parallel \mathbf{X}) < r} \underline{H}(\mathbf{Y}) \leq \hat{R}(r | \mathbf{X}).
\end{equation}
We claim that (\ref{eqn: Rhatr-lowerbound}) is equivalent to (\ref{eqn: EhatR-upperbound}). This proves the proposition.

We first show that (\ref{eqn: Rhatr-lowerbound}) implies (\ref{eqn: EhatR-upperbound}). Fix the source $\mathbf{X}$. Let $R$ be a given rate. Consider an arbitrary candidate exponent $r$ and an arbitrary source $\mathbf{Y}$. We argue that
\begin{equation}
\label{eqn: alt-condition-EhatR}
R \mbox{ is } r\mbox{-achievable and } \underline{H}(\mathbf{Y}) > R \Longrightarrow r \leq D_u(\mathbf{Y} \parallel \mathbf{X}).
\end{equation}
Taking the infimum on the right-hand side of (\ref{eqn: alt-condition-EhatR}) over $\mathbf{Y}$ with $\underline{H}(\mathbf{Y}) > R$, and then the supremum over $r$ will yield (\ref{eqn: EhatR-upperbound}).

To argue (\ref{eqn: alt-condition-EhatR}) by contraposition, we shall show that
\begin{eqnarray*}
  \lefteqn{ r > D_u(\mathbf{Y} \parallel \mathbf{X} ) } \\
    & \Longrightarrow & \mbox{either } R \mbox{ is not } r\mbox{-achievable or } \underline{H}(\mathbf{Y}) \leq R,
\end{eqnarray*}
or equivalently, we shall show that
\begin{eqnarray*}
  \lefteqn{ r > D_u(\mathbf{Y} \parallel \mathbf{X} ) \mbox{ and } \underline{H}(\mathbf{Y}) > R } \\
    & \Longrightarrow & R \mbox{ is not } r\mbox{-achievable}.
\end{eqnarray*}
But the conditions on the left-hand side imply
\[
  \sup_{\mathbf{Y}: D_u(\mathbf{Y} \parallel \mathbf{X}) < r} \underline{H}(\mathbf{Y}) > R,
\]
which together with (\ref{eqn: Rhatr-lowerbound}) yields $\hat{R}(r | \mathbf{X}) > R$, and this is the same as saying $R$ is not $r$-achievable. This completes the proof of (\ref{eqn: Rhatr-lowerbound}) $\Rightarrow$ (\ref{eqn: EhatR-upperbound}). (This direction suffices to prove Proposition~\ref{prop:sup-exponent-error-liminf}). The proof of the other direction is analogous.
\hfill \IEEEQEDclosed

To prove the upper bound in (\ref{eqn: EhatR-upperbound-alternative}), we begin with Iriyama's \cite[Eqn. (13)]{200105TIT_Iri}, which is
\[
  \sup_{\mathbf{Y}: D_u(\mathbf{Y} \parallel \mathbf{X}) < r} \{ \underline{R}(\mathbf{Y},\mathbf{X})-D_u(\mathbf{Y}\parallel \mathbf{X}) \} \leq  \hat{R}(r|\mathbf{X}),
\]
instead of (\ref{eqn: Rhatr-lowerbound}). The rest of the proof is completely analogous to the proof of Proposition \ref{prop:sup-exponent-error-liminf}.

\section{Proof of Proposition \ref{prop:inf-rate-correct}}
\label{subsec:inf-rate-correct-proof}
We use the following notations in this proof. For each  $\mathbf{B}=(B_n: n \in \mathbb{N})$ define
\[
|\mathbf{B}|:=\limsup_{n \rightarrow \infty}\frac{1}{n}\ln |B_n|
\]
and
\[
S(\mathbf{Y}):=\left\{\mathbf{B}: \lim_{n \rightarrow \infty} P_{Y^n}(B_n)=1\right \}.
\]
Note that $\mathbf{B} \in S(\mathbf{Y}) \Leftrightarrow \mathbf{Y} \in \mathcal{M}(\mathbf{B})$. We will first prove (\ref{eqn:correct-rate-expression-liminf}). Define a set
\begin{eqnarray}
\lefteqn {\mathcal{B}(r,\rho|\mathbf{X})=\bigg \{\mathbf{B}:=(B_n: n \in \mathbb{N}):} \nonumber\\
\label{eqn:VariationalSet}
&&\hspace*{.2in}(1+\rho) \liminf_{n\rightarrow \infty}\frac{1}{n}\ln \sum_{x^n \in
B_n}P_{X^n}^{\frac{1}{1+\rho}}(x^n) \geq r\bigg\}.
\end{eqnarray}
Then, by definition,
\begin{equation}
\label{eqn:Rate-definition-renyi}
R^{*}(r,\rho|\mathbf{X})=\inf \left \{|\mathbf{B}|: \mathbf{B} \in \mathcal{B}(r,\rho|\mathbf{X})\right \}.
\end{equation}
Fix a $\mathbf{B} \in \mathcal{B}(r,\rho|\mathbf{X})$. Proposition \ref{prop: limit-variational-formula} then implies
\begin{eqnarray*}
\lefteqn{ (1+\rho) \liminf_{n\rightarrow \infty}\frac{1}{n} \ln \sum_{x^n \in B_n}P_{X^n}^{\frac{1}{1+\rho}}(x^n) } \\
 & & \hspace*{1in}  = ~ \max_{\mathbf{Y}:\mathbf{B}\in S(\mathbf{Y})}E_l(\mathbf{Y},\mathbf{X},\rho).
\end{eqnarray*}
We can therefore conclude using (\ref{eqn:VariationalSet}) that the following set equivalence holds:
\begin{equation}
\label{eqn:set-equivalence}
\mathcal{B}(r,\rho|\mathbf{X})=\bigcup_{E_l(\mathbf{Y},\mathbf{X},\rho)\geq r} S(\mathbf{Y}).
\end{equation}
From (\ref{eqn:Rate-definition-renyi}) and (\ref{eqn:set-equivalence}) we get
\begin{eqnarray}
R^{*}(r,\rho|\mathbf{X})&=&\inf\left\{|\mathbf{B}|:\mathbf{B} \in \bigcup_{E_l(\mathbf{Y},\mathbf{X},\rho)\geq r} S(\mathbf{Y})\right\}\nonumber\\
&=&\inf_{\mathbf{Y}}\{|\mathbf{B}|:E_l(\mathbf{Y},\mathbf{X},\rho)\geq r,\mathbf{B}\in S(\mathbf{Y})\}\nonumber\\
\label{eqn:Han-Verdu-entropy}
&=&\inf_{\mathbf{Y}:E_l(\mathbf{Y},\mathbf{X},\rho)\geq r}\overline{H}(\mathbf{Y}), \nonumber
\end{eqnarray}
where last equality follows because
\[
\overline{H}(\mathbf{Y})=\inf \left \{|\mathbf{B}|: \mathbf{B} \in S(\mathbf{Y})\right \}
\] as proved by Han \& Verd\'{u} \cite{199305TIT_HanVer}. This proves (\ref{eqn:correct-rate-expression-liminf}).

We now prove (\ref{eqn:correct-exponent-expression-liminf}). We first show that if $R$ is $(r,\rho)$-admissible then $r\leq \sup_{\overline{H}(\mathbf{Y})\leq R}{E}_l(\mathbf{Y},\mathbf{X},\rho)$.\\
\noindent Since $R$ is  $(r,\rho)$-admissible, definition of $R^*(r, \rho|\mathbf{X})$ and (\ref{eqn:correct-rate-expression-liminf}) imply
 \[
R\geq R^*(r,\rho|\mathbf{X})= \inf_{\mathbf{Y}:E_l(\mathbf{Y},\mathbf{X},\rho)\geq r}\overline{H}(\mathbf{Y}),
\]
i.e., for all $\delta >0$ there exists a $\hat{\mathbf{Y}}$ such that
\[
E_l(\hat{\mathbf{Y}},\mathbf{X},\rho)\geq r~~ \textnormal{and} ~~\overline{H}(\hat{\mathbf{Y}})<R+\delta,
\] which further implies that
\[
r \leq \sup_{\overline{H}(\mathbf{Y})< R+\delta}{E}_l(\mathbf{Y},\mathbf{X},\rho).
\]
Since $\delta$ was arbitrary, letting $\delta \downarrow 0$ yields
\[
r\leq \sup_{\overline{H}(\mathbf{Y})\leq R}{E}_l(\mathbf{Y},\mathbf{X},\rho),
\] and the converse part is proved.

For the direct part it is sufficient to show that given $\rho$, any $R$ with
\[
r:=\sup_{\overline{H}(\mathbf{Y})\leq R}{E}_l(\mathbf{Y},\mathbf{X},\rho),
\]
is $(r,\rho)$-admissible. By choice of $r$, for all $\delta>0$, there exists a $\hat{\mathbf{Y}}$ such that
\[E_l(\hat{\mathbf{Y}},\mathbf{X},\rho)> r -\delta~~\textnormal{and}~~\overline{H}(\hat{\mathbf{Y}})\leq R.\]
This implies that
\[
\inf_{E_l(\mathbf{Y},\mathbf{X},\rho) > r -\delta}\overline{H}(\mathbf{Y})\leq R.
\]
Since $\delta$ was arbitrary, let $\delta \downarrow 0$ and use (\ref{eqn:correct-rate-expression-liminf}) to get
\[
R \geq \inf_{E_l(\mathbf{Y},\mathbf{X},\rho) \geq r }\overline{H}(\mathbf{Y})=R^*(r,\rho|\mathbf{X}),
\]
i.e., is $(r,\rho)$-admissible. This completes the proof.
\hfill \IEEEQEDclosed

\bibliographystyle{IEEEtran}
{
\bibliography{IEEEabrv,wisl}
}

\end{document}